\RequirePackage{fix-cm}
\documentclass[smallextended]{svjour3}       % onecolumn (second format)
\smartqed  % flush right qed marks, e.g. at end of proof
\usepackage{graphicx}

\usepackage[mathlines,switch]{lineno}
\usepackage[justification=centering]{caption}
\usepackage{array,multirow}
\usepackage{subcaption}
\usepackage{url}
\usepackage{balance}

\usepackage{tablefootnote}
\usepackage{threeparttable}
\usepackage{amssymb}

\usepackage{listings}
\usepackage{color}
 \usepackage[T1]{fontenc}
 
\definecolor{codegreen}{rgb}{0,0.6,0}
\definecolor{codegray}{rgb}{0.5,0.5,0.5}
\definecolor{codepurple}{rgb}{0.58,0,0.82}
\definecolor{backcolour}{rgb}{0.97, 0.97, 0.97}
\lstdefinestyle{mystyle}{
    backgroundcolor=\color{backcolour},   
    commentstyle=\color{codegreen},
    keywordstyle=\color{magenta},
    numberstyle=\tiny\color{codegray},
    stringstyle=\color{codepurple},
    basicstyle=\footnotesize,
    breakatwhitespace=false,         
    breaklines=true,                 
    captionpos=b,                    
    keepspaces=true,                 
    numbers=left,                    
    numbersep=5pt,                  
    showspaces=false,                
    showstringspaces=false,
    showtabs=false,                  
    tabsize=2,
    framesep=10pt
}
 
\begin{document}

\title{A Machine Learning Based Framework for Code Clone Validation%\thanks{Grants or other notes
}


\author{Golam Mostaeen \and Banani Roy \and Chanchal Roy
\and
Kevin Schneider
\and
Jeffrey Svajlenko
}


\institute{F. Author \at
              Tel.: +000-00-000000\\
              Fax: +000-00-000000\\
              \email{anonymous@example.com}           %  \\
           \and
           S. Author \at
              Tel.: +000-00-000000\\
              Fax: +000-00-000000\\
              \email{anonymous@example.com}           %  \\
}


\titlerunning{Clone Validation using Machine Learning}
\authorrunning{Journal of Systems and Software}

\maketitle

\setcounter{tocdepth}{3}
%\newpage
%\tableofcontents

%\newpage

\begin{abstract}
A code clone  is a pair of code fragments, within or between software systems that are similar. Since code clones often negatively impact the maintainability of a software system, several code clone detection techniques and tools have been proposed and studied over the last decade. However, the clone detection tools are not always perfect and their clone detection reports often contain a number of false positives or irrelevant clones from specific project management or user perspective. To detect all possible similar source code patterns in general, the clone detection tools work on the syntax level while lacking user-specific preferences. This often means the clones must be manually inspected before analysis in order to remove those false positives from consideration. This manual clone validation effort is very time-consuming and often error-prone, in particular for large-scale clone detection. In this paper, we propose a machine learning approach for automating the validation process. First, a training dataset is built by taking code clones from several clone detection tools for different subject systems and then manually validating those clones. Second, several features are extracted from those clones to train the machine learning model by the proposed approach. The trained algorithm is then used to automatically validate clones without human inspection. Thus the proposed approach can be used to remove the false positive clones from the detection results, automatically evaluate the precision of any clone detectors for any given set of datasets, evaluate existing clone benchmark datasets, or even be used to build new clone benchmarks and datasets with minimum effort. In an experiment with clones detected by several clone detectors in several different software systems, we found our approach has an accuracy of up to 87.4\% when compared against the manual validation by multiple expert judges. The proposed method also shows better results in several comparative studies with the existing related approaches for clone classification.
\end{abstract}

\section{Introduction}
\label{intro}
Copying and reusing certain pieces of existing code directly
or with alteration into another location is a common
programming practice in a software development life cycle \cite{roy2007survey}. Such copy/paste practice results in similar pieces of code fragments in a system, called code clones.
Researchers agree upon four primary clone types  \cite{roy2007survey}: Type-1
clones are syntactically identical code fragments, regardless of
the presentation style, comments, and white spaces. Type-2
clones are copy and pasted code where identifier names and
types have been changed. Type-3 clones are modified copies of
the original code with statement-level changes (e.g., additions of new statements, or deletions and modifications of existing ones). Type-4 clones are syntactically dissimilar code fragments that implement the same or similar functionality. Some of the recent research shows that on average around 7\%
to 23\% of total code of a software system is duplicated or
cloned from one location to another \cite{baker1995finding}, \cite{kapser2006supporting}\cite{MondalICPC2011Maintenance}. Although code
cloning is often done intentionally to accelerate the development
process and also not all code clones are harmful \cite{kapser2006cloning}, the
existence of some of them can inflate software maintenance
costs as clones are one of the major causes of creation and
propagation of software bugs throughout the system \cite{juergens2009code}, \cite{gode2011frequency}, \cite{bettenburg2009empirical}, \cite{Saha_2013_Type3Evolution}. For
example, it becomes very difficult to carry out consistent
changes to all the cloned code fragments throughout the
software system. These inconsistent changes to the corresponding
duplicated code fragments are often responsible for the creation of
new software bugs \cite{mondal2017bug}. In addition to the creation of new bugs, code
cloning also becomes one of the main reasons for bug
propagation when programmers copy-and-paste a buggy code
fragment throughout the software system for implementing
similar functionalities \cite{mondal2017cloned}, \cite{juergens2009code}. Detection of such code clones can,
 therefore, accelerate the maintenance tasks of any software
systems remarkably \cite{juergens2009code}. Besides, exploiting the similarities of the
detected code clones also helps one better understand and improve
the overall software design \cite{higo2002software}, \cite{MondalWCRE2014}, \cite{MondalSCAM2014}, \cite{ZibranACMSAC2012}.\\
\indent At least 70 Clone Detection Tools and techniques have been proposed and
developed to automate the clone detection process, for the 
extensive research in this specific area over the last decade \cite{kamiya2002ccfinder}, \cite{baker1993program}, \cite{duala2007tracking}, \cite{bellon2007comparison}, \cite{roy2008nicad}, \cite{jiang2007deckard}, \cite{tairas2006phoenix}. These tools return a list of possible code
clone pairs or classes available in a given software system. Except for Type 1,
the other types of code clones (Type 2, 3 and 4) undergo
different changes over time and can get too complicated to be
detected with a simple string matching algorithm by a clone detection tool. For example, the
identifiers or functions names may be changed, some code
fragments may be added, modified or removed, a portion of the
code clones might undergo several other syntactical changes or
even the complete implementations might be changed for the
same functionalities in any other locations etc. All these
modifications over time make the searching problem much
more complicated. In order to handle those complex source code
structures while still detecting all possible code clone pairs, the
tools undergo a lot of generalization of the original source
codes like pretty-printing \cite{roy2008nicad}, normalization of the identifiers
\cite{kamiya2002ccfinder}, \cite{roy2008nicad}, forming syntax tree \cite{koschke2006clone} of the code fragments and so
on just to name a few.\\
\indent As a result of this complex searching problem and necessary
generalization or normalization of the source code, the clone
detection algorithms often report false positive clones. These are
pairs of code fragments that are not similar or possibly are only
coincidentally similar or are otherwise considered not a valid
clone by the user \cite{jiang2007framework}. Besides, some research shows that the
definition of true positive code clones, especially in case of Type
3 and Type 4 clones, are subjective and might also be different for
different users or software systems \cite{keivanloo2015threshold}, \cite{charpentier2015empirical}, \cite{yang2015classification}. For example, Yang et. al. \cite{yang2015classification} conducted a survey where several users were provided the same clone sets for validation detected by clone detection tools. The study reported significant variations among the users in validating the same clone sets (e.g. for the same provided clone sets, the number of decided true positive code clones varied within a range of 4.76\% to 23.81\% for different users).  For these reasons,
programmers often need to manually validate if the results of a clone detection are a true
clone or not before using this information for the given specific
scenarios like source code refactoring or other software
maintenance tasks. Such a manual validation process becomes a hindering factor even for a medium sized software system.
Because in that case programmers often find it challenging to
extract the actual true positive clones they are looking for from
those large set of reported possible code clone pairs by clone
detection tools. For example, some previous research shows that
JDK 1.4.2 contains 204 K LoC reported code clone which is 8\% of
the total lines of code \cite{jiang2007framework}, \cite{tairas2009information}. 15\% of the total lines of code of the
Linux kernel has been reported as code clone which is 122 K LoC
\cite{li2006cp}. Both of the above scenarios on the number of reported possible
code clones by clone detection tools illustrate the huge amount
of manual validation work necessary before using the code clone information. Besides, the clone
detection algorithms of the tools usually work in general
irrespective of the specific system requirements or user
preferences. Thus, in the best case even if a tool returns only true positive clones, many of those clones might not be relevant to the tasks at hand of the programmers or engineers (e.g., not suitable for refactoring) \cite{jiang2007framework}. Mining those code clones of
interest from the tool generated report is often a time-consuming task and thus reduces the usability of code clone
detection tools. The scenario gets even worse with the
increase of software project in size. Studies also show the subjective nature of detected clones requiring manual validation before usage. For example, in the conducted user study of  Yang et al \cite{yang2015classification}  involving 105 detected clone pairs and multiple expert judges were asked for manual validation. Among the set of 105 detected clone pairs, the independent manual judges showed significant differences in the validation patterns (i.e., as presented above). For example, one of the user judged 5 of the clone pairs as false clone, while other users judged  24, 23 and 25 of the clone pairs as false clones respectively.\\
\indent In this paper we propose a machine learning based approach for predicting the user code clone validation patterns. The proposed method works on top of any code clone detection tools for classifying the reported clones as per user preferences. The automatic validation process for a user, thus can accelerate the overall process of code clone management and helps faster acquiring of required information out of the clones in comparison to the time-consuming manual validation process. We studied performance and result qualities of different machine learning algorithms in validating the detected clones. We also extend the proposed method with a cloud-based architecture to ensure the compatibility of the proposed method with any of the existing clone detection tools. We got promising results from our several studies with different experimental setups for the clone validation. The proposed method also showed the better result in a comparison study with related existing works for code clone validation.\\
\indent This study is aimed to answer the following 2 research
questions:
\begin{itemize} %MODIFY: RQ1 focus on user specific val
    \item \textbf{RQ 1}: Can the manual code clone validation process be assisted via machine learning?
    
    %\item \textbf{RQ 2} : Can the machine learning technique predict a programmer’s validation behaviour to improve itself over time and experiences?
    
    \item \textbf{RQ 2}: Does the proposed machine-learning based validation method work across different clone types and clone detection tools?
\end{itemize}
\indent Our work makes three main contributions. \textit{First}, we propose a cloud-based architecture for automatic code clone validation using machine learning. We also present an implementation of the working version of the tool as a proof of concept of the proposed method. The open-source tool is available at GitHub \footnote{https://github.com/pseudoPixels/CloneCognition} for further extension and contribution to the research domain. \textit{Second}, we studied the data distribution for the clone classification problem with several extracted features. Our findings on these feature sets and data distribution analysis can help better understand the clone classification problem and thus adds the possibility of result improvement in this research area. \textit{Third}, we conducted a detailed comparative study with 11 different machine learning algorithms for the clone classification. To the best of our knowledge, no previous studies were done that focused on a comparative study of different machine learning algorithms for clone classification problem. Our observations on the strengths and weaknesses of several machine learning algorithms on clone classification can contribute to future research in this area for further improvement of the learning model.\\
\indent The rest of this paper is organized as follows: Section \ref{sec:rel_works} provides a discussion of related works on this specific research area. Section \ref{sec:proposed_method}, contains a discussion about the proposed method. We then present the data distribution analysis and comparative analysis of multiple machine learning algorithms in Section \ref{sec:data_distribution} and Section \ref{subsec:mult_ml_perform} respectively. Section \ref{sec:exp} discusses the several experiments we conducted with proposed and existing related methods. The result discussion and possible threats to validity are presented in Section \ref{sec:res_discusion} and Section \ref{sec:threats_to_validity} respectively. Finally, in Section \ref{sec:conclusion}, we discuss our conclusion and future work.
%%%%%%%%%%%%%%%%%%%%%%%%%%%%%%%%%%%%%%%%%%%%%%%%%%%%%%%%%%%%%%%%%%%%%
%                   WRITE CONTRIBUTION ABOVE
%%%%%%%%%%%%%%%%%%%%%%%%%%%%%%%%%%%%%%%%%%%%%%%%%%%%%%%%%%%%%%%%%%%%%

\section{Related Works}\label{sec:rel_works}
In this section, we introduce existing research works which targeted the clone classification problem and closely related to our proposed method for automatic clone validation.\\  %systems for user specific code clone validation along with related works and then we present available scopes of improvements and motivations for our work.
%\subsection{Traditional Approaches for Code Clone Classification}
\indent Though several methods and techniques have been proposed
over the past few years for maintenance, organization or classification of code
clones, very few of them recently focused on aiding the huge manual user-specific
validation task of the reported code clones. Yang et. al. \cite{yang2015classification} studied the similar problem for user-specific code clone classification in their work  - FICA. The user-specific clone classification in FICA is done by token sequence similarity analysis using \lq Term-Frequency - Inverse Document Frequency\rq - (TF-IDF) vector. For training some of the reported code clone pairs from clone detection tools are manually marked as True or False positive clones by the users. So, the whole training set $M$, is divided into two sets of clones- $M_t$ and $M_f$ such that $M_t\cap M_f=\emptyset$ and $M_t\cup M_f=M$. Tokens are then extracted from both the clone sets to produce an n-gram (considered N=3 for the study) of token sequences. Defining term $t$ as an n-gram of token sequence and document $d$ as a clone set, the Term Frequency (TF) are calculated as  Eq. 1.
\begin{equation}
TF(t,d)= \frac{t : t \in d}{|d|}
\end{equation}
Similarly defining documents $D$, as all the clone sets of a project or different software systems of considerations, the Inverse Document Frequency (IDF) and TF-IDF vector are calculated using Eq. 2 and Eq. 3 respectively. 
\begin{equation}
IDF_D(t)= \log\frac{|D|}{1 + | d\in D : t\in d|}
\end{equation}
\begin{equation}
\overrightarrow{TF\textnormal{-}IDF_D(d)}= [TF\textnormal{-}IDF_D(t,d) : \forall t \in d]
\end{equation}
Using TF-IDF for the two clone sets $a$ and $b$, cosine similarity $CosSim_D(a,b)$ are then calculated for a set of documents $D$. The probability score for an unmarked clone set $c$, for being in clone set $M_t$ or $M_f$ are then calculated as in Eq. 4. 
\begin{equation}
  P_{M_x}(c) = \frac{\sum\limits_{\forall m \in M_x} CosSim_{M_x} (c, m) . w(m)}{\sum\limits_{\forall m \in M_x} w(m)}
\end{equation}
Here, $w(m)$ is the assigned weight in the range $[0,1]$, in the case of numerical range marking of the clones. However, in case the of boolean marking (i.e. either $M_t$ or $M_f$) of clones, $w(m)$ is uniformly set to 1. Besides, for the improvement of the model, FICA optionally takes user feedback iteratively over time to populate the training set $M_t$ and $M_f$. 
As FICA learns user-specific validation completely based on token sequence, the validation accuracy gets significantly lower as the target clone goes beyond Type 2 as also noticeable from the study. \\ %MODIFY: add a few more lines...
%\subsection{Related Works}
\indent In recent years a number of research studies have been done for code clone detection tools' reported clone classification  or comprehension. For example, Tairas et. al. \cite{tairas2009information} broadly classified the 
existing code clone comprehension techniques into two categories. The first category of the techniques does the
classification of the detected code clones based on
certain properties: location of the clones with respect to one
another in the hierarchy files and directories \cite{kapser2004aiding}, type
similarities (all possible Type 1 clones grouped together and so
on for Type 2, 3 and 4) of the detected clones \cite{bellon2007comparison}, and Latent Semantic
Indexing (LSI) on the identifiers of clones \cite{tairas2009information}.
Besides, 
some machine learning algorithms have also been applied to
group the detected clones: token sequence similarities of the
clones have been analyzed to categorize them \cite{yang2015classification}, applying
unsupervised machine learning algorithm to create clone
clusters \cite{svajlenko2016machine}. On the other hand, the second category
of clone comprehension techniques works based on their visual
representation: scatter plot of the code clones \cite{kamiya2002ccfinder}, an aspect
browser-like view \cite{tairas2009information}, and hierarchical graphs of detected clones
\cite{jiang2007deckard}. These classification and visualization techniques
can make the organization and maintenance of code clones are much easier.
%\subsection{Motivation} %MODIFY: shorten a few lines. focus on user specific validation.
We can notice that the total number of clones to be
manually analyzed for validation still remains the same. The
overall result of such code clone comprehension techniques can
be improved significantly by adding an automatic validation
process that uses a machine learning approach to learn to validate
according to the specific system and user over time.\\
\indent Besides, researchers often find it challenging to evaluate any tools or
techniques on clone detection due to the lack of enough
validated code clone benchmark. Because building such
benchmarks often contain possible threats to the validity due to
unavoidable human errors and need a huge amount of manual
validation work. For example, Bellon et al. \cite{bellon2007comparison} created one such
benchmark by validating 2\% of the union of six clone detectors
for eight subject systems that required 77 hours of manual
efforts, Svajlenko et al. \cite{svajlenko2014towards} created a benchmark of true positive
clones that also reports hours of manual validation efforts. So, the trained machine learning model can be used to aid in the creation of user specific validated clone sets. \\
\indent The proposed method works as a layer on top of the reported
possible code clone pairs generated by existing clone detection
tools. Initially some user validated clones are fed to the system
for learning the validation behavior of the specific system or
programmer. The proposed method extracts several features right
from the reported code clone pairs of source code fragments. Once the training phase with validated code
clone pair is completed, the system is given unknown or test code
clone pairs to validate. The proposed system extracts the exact
same features from the test code clone pairs and feeds them to
the trained machine learning algorithm where it gives the
validation response. The programmer's feedback result can optionally be
given to the system to update and improve its prediction rule.
This gives the proposed method an opportunity to improve and 
learn the programmer's preferences even better with time and
experience. Besides, the proposed method can optionally be
tuned to control the validated result based on the importance of
the code clone pairs including in the IDE-based management of clones \cite{ZibfranIWSC2011IDE} or when searching clones in the IDE \cite{Zibran_2012_CloneSearch}. That is a programmer can choose to get all
the code clone pairs reported by a clone detection tools or only
those code clone pairs having some particular importance as per
their preferences. This gives the programmer flexibility on
selecting the number of validated clones for given
scenarios.

\section{Proposed Approach} \label{sec:proposed_method}
In this section, we discuss the proposed method for the clone classification problem. In Section \ref{subsec:overall_workflow}, we first present the high-level workflow of the proposed method for training the machine learning model. Section \ref{subsec:feature_extraction} to Section \ref{subsec:improving_ml_models_w_feedback}, contain the discussions detailing some of the significant workflow steps. And finally, we present the extended cloud-based architecture of the proposed method in Section \ref{subsec:cloud_arch}. 

\subsection{Overall Workflow of the Proposed Method}\label{subsec:overall_workflow}
 Figure \ref{fig_proposed_workflow}, shows a high-level workflow (which has been extended for the cloud-based architecture as discussed in Section \ref{subsec:cloud_arch} in details) of  the proposed method. The proposed method uses machine learning models for predicting the user-specific code clone validation. The models are first trained based on manually validated code clone sets from the corresponding users. The trained models are then used for improving the reported code clones from clone detection tools by predicting the user-specific validation patterns. The workflow steps can be listed in sequence like the following:
 \begin{figure}[!htbp]
\centerline{\includegraphics[scale=0.5]{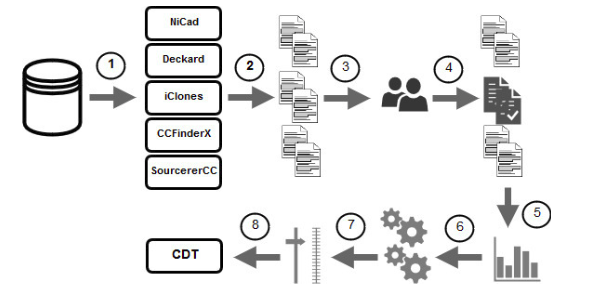}}
\caption{Workflow of the proposed method}
\label{fig_proposed_workflow}
\end{figure}
 \begin{enumerate}
  \item In this step, source code from Codebase are supplied to any of the existing clone detection tools.
  \item The detected code clones from the corresponding clone detection tools are collected in Step 2. As the proposed method works with any of the existing clone detection tools, clones from multiple tools can be combined optionally for further generality of the training set.
  \item The reported code clones from clone detection tools are provided to the user for manual validation.
  \item User marks the code clones as true positive or false positive in Step 4 (details in Section \ref{subsec:feature_extraction}). The user-specific manual clone validation results are stored in a database for use as a training set of machine learning models.
  \item Several features are extracted from the marked code clone pairs for training the machine learning models. The existing related research works (for example FICA \cite{yang2015classification}), used only simple token sequences as features for training the machine learning model and thus failed to predict the validation successfully beyond Type 2 clones. To improve the classification results and to target clones beyond Type 2, we considered calculating clone similarity with several levels of structural pre-processing and normalization. The details of the feature extraction methodologies have been presented in Section \ref{subsec:feature_extraction}.
  \item The extracted features are used to build the feature vector for clone classification. Feature vectors for the corresponding manual validation clone classes are used for training machine learning model in Step 6.
  \item Next, in Step 7, the trained machine learning model is used for predicting the clone validation pattern for the unknown or test sets. The machine learning model at this stage returns the probability score (of being true or false positive) for the given corresponding code clone pairs. 
  \item Finally, in this step, the classified result is sent back to the clone detection tools. The classification result can be tuned based on user preference (of probability score) for the final result. The system can take user feedback based on the classified clones from the user for repeating the cycle of supervised learning, thus improving the validation result over time.
\end{enumerate}

\subsection{Clone Labeling and Feature Extraction}\label{subsec:feature_extraction}
The reported code clones from the clone detection tools are provided to the user for manual validation (Step 3, Figure \ref{fig_proposed_workflow}). The corresponding user validation results are stored in a database which is later used for training the machine learning model. Reported clones from clone detection tools are used to create a clone database, $K$. Clones from $K$, are manually marked as true or false positive by the user. Reported code clones are thus grouped into two disjoint sets $K_t$ and $K_f$ - representing true positive and false positive clone groups respectively such that, $K=K_t \cup K_f$ and $K_f\cap K_t = \emptyset$. $K_t$ and $K_f$ are used for training the machine learning models.\\
\indent As machine learning models learn to map the input feature sets to the corresponding class label, it is important to select appropriate features for the given classification problem. For example, Yang et. al. \cite{yang2015classification} targeted Type 2 clones in a similar study of the code clone classification problem and hence used simple token sequences as features for using in the classification algorithm. As we intended to enhance  the classification performance (so that it works efficiently with more diverse types of clones and also shows better prediction score), in addition to improving the whole classification workflow, we also focused on extracting more informative features. Most of the selected feature extraction undergoes two main steps: i) Pre-processing and source code transformation and ii) Similarity extraction from the code clone pairs.\\
\indent Pre-processing, like pretty-printing and comment removal ensures consistent structures for matching and similarity extraction of similar source code pairs (for example Type 1 clones). Extracting similarity features between the two code clone fragments after this step (i.e. comment removal followed by pretty-printing) gives us the information about how a user sees the code clones for validation. Thus at this point, the extracted features represent mainly Type 1 similarity between the target code clone fragments. In addition to that, different source transformations like consistent normalization of literals or consistent renaming of identifiers are applied to consider the possible changes between the code clone pairs (i.e. for Type 2 and Type 3 clones). For example, Listing 1 and Listing 2, show the code fragments of one of the detected clones from Weka \cite{weka} software system, that needs to be validated (comparatively simpler and straight forward code clone pair used for discussion).  
\begin{lstlisting}[language=Java, caption=Sample Code Clone (Fragment 1)]
try {
    if (args.length == 0) {
	throw new Exception(
	    "The first argument must be the class name of a kernel");
    }
    String associator = args[0];
    args[0] = ">";
    System.out.println(evaluate(associator, args));
}
\end{lstlisting}
\begin{lstlisting}[language=Java, caption=Sample Code Clone (Fragment 2)]
try {
    if (args.length == 0) {
	throw new Exception(
	"The first argument must be the name of a " 
	+ "clusterer");
    }
    args[0] = "?";
    Clusterer newClusterer = AbstractClusterer.forName(ClustererString, null);//object from abstract clusterer
    System.out.println(evaluateClusterer(newClusterer, args));
}
\end{lstlisting}

Though the code clone pairs exhibit much structural similarity, calculating similarity directly based on original source code pairs have higher probability of introducing noise (from the perspective of Type 2 or Type 3 clones) due to strict consideration of the modifications of literals and identifiers. So, we also applied different source code transformations before calculating clone similarity for extracting possible Type 2 or Type 3 information. For example,

\begin{lstlisting}[language=Java, caption=Pre-processed and Transformed Code Clone (Fragment 1)]
try {
    if (X.X == 0) {
	throw new X(
	    "string");
    }
    X X = X[0];
    X[0] = "string";
    X.X.X(X(X, X));
}
\end{lstlisting}
\begin{lstlisting}[language=Java, caption=Pre-processed and Transformed Code Clone (Fragment 2)]
try {
    if (X.X == 0) {
	throw new X(
	"string" 
	+ "string");
    }
    X[0] = "string";
    X X = X.X(X, null);
    X.X.X(X(X, X));
}
\end{lstlisting}
Listing 3 and Listing 4, show the transformed clone fragments from Listing 1 and Listing 2 respectively, after the first  blind renaming of identifiers and then applying consistent normalization of literals. For example, after the blind renaming of the identifiers, all different identifiers take a common name, X for the structural comparison. Similarly, all the different string literals were transformed to - "string" after consistent normalization of the literals as shown in Listing 3 and Listing 4. These transformations allow the corresponding modifications of literals and identifiers and thus provides similarity feature information for Type 2 and Type 3 code clones. We used TXL \cite{cordy1991txl} for different source transformations.\\
\indent After applying different pre-processing and transformation for different types of features, we then analyze the differences between the code clone fragments (i.e. the output of the previous steps). Prior to calculating numerical similarity values between the clone fragments, we find out the minimal changes required to transform from one clone fragment to another. For example, Listing 
\begin{lstlisting}[caption=Difference between the code clone fragments]
4c4,5
< 	    "string");
---
> 	"string" 
> 	+ "string");
6d6
<     X X = X[0];
7a8
>     X X = X.X(X, null);
10,12d10
< 
< 
<
\end{lstlisting}
5, shows the minimal changes or operations required for transforming code clone fragment 1 (i.e. Listing 3) to code clone fragment 2 (i.e. Listing 4). We used the \textit{Unix Diff} utility for the purpose, that calculates the minimum set of insert and delete operations required for converting one file to another. For example, in Listing 5, $<$ and $>$ signs - represent delete operation, $\mathcal{O}_d$ and insert operation $\mathcal{O}_i$ respectively, that we need to apply on first clone fragment for the required transformation. For example, the first conflict of the transformed clone fragments in Listing 3 with Listing 4 is at line 4. The minimum operations needed to resolve the difference is one delete operation, $\mathcal{O}_d$ of the original line at 4, followed by two insert operations, $\mathcal{O}_i$ of line 4 and 5 from Listing 4. The corresponding change operations have been represented as $4c4,5$ in Listing 5. We then calculate the similarity value between the two code clone fragments $f_1$ and $f_2$ as, $\xi(f_1, f_2) = 1 - max(C(\mathcal{O}_d)/|f_1|, C(\mathcal{O}_i)/|f_2|)$, where $C(\mathcal{O})$ and $|f|$, represent the count of the corresponding change operation and length of the corresponding code clone fragment respectively. The fragment similarity thus falls in the range of [0,1]. As the number of such differences between the two code clone fragments increases, the code clone fragments similarity measure tends towards zero. On the other hand, the fragment similarity is calculated as 1, in case the clone fragments are exactly similar with no further required changes (i.e. $C(\mathcal{O}_d) = C(\mathcal{O}_i) = 0$).\\
\indent We also used several other features to get more structural information about the two code clone fragments. We tried to mimic several manual validation patterns as per our obtained experiences on manual code clone validation of users. For example, our intuition was if the code clone fragments are
significantly different in size, a validator may be more likely to
mark them as false positive. The corresponding code clone fragment sizes $\alpha$ and
$\beta$, were calculated as  respectively. The difference $|\alpha-\beta|$, provides information about the variation of 
fragment sizes. Smaller difference values represents more likelihood of being validated as true positive code clone than
that of comparatively higher difference values and thus was considered as one possible feature for the clone classification problem.
However, for a clone that is small versus a clone that is
large might have different consideration. For example, for a relatively larger code clone fragment pair, it is possible to have more variance in difference than that of smaller code clone fragments pair. So, to mitigate this possible bias we also considered 
the average size of the code clones $(\alpha+\beta)/2$. That average value
captures the size of the clones and difference captures if
the code fragments are rather mismatched in size.\\
\indent Please note that some of the popular code clone detection tools use source transformations like consistent renaming of identifiers,  or normalization of literals, as part of their workflow for code clone detection. A few such clone detection tools like  NiCad \cite{roy2008nicad} and CCFinder \cite{kamiya2002ccfinder} are also well known in the research area for their performance on clone detection. So, with this motivation, for a subset of the features we also carried out similar transformations before calculating the clone fragment similarities as discussed above. However, to the best of our knowledge no previous works on clone validation used a similar feature set, hence before finalizing the feature selection for building the machine learning models, we conducted several studies on data distribution with the extracted features. For a given feature from the feature set, we tried to find out its class separability for the two classes and its overall contribution score for the classification. Some of the features showed higher contribution score while a few a of them showed comparatively low contribution score or class separability. The feature study provided us the information about higher contributing features while removing a few of them that more or less exhibit as possible noise for the clone classification problem. Section \ref{sec:data_distribution}, discusses our findings in details about the feature sets for the classification of the code clones.
 
%
%In the next step, we then extracted previously analyzed features from those manually validated clone pairs for training. TXL \cite{cordy1991txl} was used for normalization and several other pre-processing step on the source code before finding the features. We used Java then on those pre-processed and normalized source code to extract features. We selected several features that shows better result based on
%the study of data distribution. We selected 8 such features (Table \ref{table:selected_features}): Syntactical similarity both of line and token
%after Type-1, Type-2 and Type-3 normalization separately,
%Number of unmatched braces for the code fragments and
%Number of intersected functions between these possible code
%clone pairs.

\subsection{Training Machine Learning Models for Clone Classification}\label{subsec:training_ml_models}
As we have presented the workflow of the proposed method in the above discussion, it uses a supervised machine learning algorithm for learning the classification pattern of the user-specific clone validation (i.e. in Step 6). The supervised classification algorithm will be trained on the manually validated datatset $D=\{(\mathbf{x}_1, \mathbf{y}_1), (\mathbf{x}_2, \mathbf{y}_2) ... (\mathbf{x}_m, \mathbf{y}_m)\}$, for $\mathbf{x}_i\in \mathbb{R}^n$ and $\mathbf{y}_i\in \mathbb{R}^l$, where $n$ and $l$ represent the extracted clone feature set and clone validation labels respectively. The machine learning algorithm is then trained on dataset $D$, to learn a function $f$, such that $f$ can map from $\mathbb{R}^n$ to $\mathbb{R}^l$, representing the class probability for being true or false positive for the given pairs of code clones.\\
\indent We investigated the classification performance using different machine learning algorithms as to the best of our knowledge, we could not find any other previous research works that directly focused on user-specific clone
validation using such extracted clone features to target validation of all 3 different types of clones. The most relevant works we found used some sequence matching algorithms instead (for example, TF-IDF, token sequence matching) and failed to validate beyond Type 2 clones (for example FICA \cite{yang2015classification}). We studied the performance of multiple machine learning classification algorithms, for example, Random Forest, Naive Bayes Classifier, Decision Table, Backpropagation Neural Networks, Deep Learning. In the comparative study of such 13 different machine learning algorithms, we got accuracies within a range of 76\% to 87\% for clone classification. Backpropagation Neural Network resulted in maximum performance in comparison to other classification models with an accuracy of 87.4\%. So, we used the Backpropagation Neural Network as the machine learning model for the experiments of the proposed method (as discussed in Section \ref{sec:exp}). Our findings and results of the comparative study of multiple machine learning classification algorithms have been discussed in detail in Section \ref{subsec:mult_ml_perform}. 

\subsection{Prediction Decision Configuration}\label{subsec:prediction_decision_config}
The machine learning models classify the test code clone pairs using the extracted feature vector $\mathbf{x}_t$. Probabilistic classifiers learn a function $f$, such that $f(\mathbf{x_t})$, assigns probability values, $\hat{\mathbf{y}_t}$ for the two classes, where $Pr[\mathbf{y}_t = (1,0)]$, represents the probability of belonging to true positive clone class. In the proposed method, users can set the decision threshold
$\gamma[0, 1]$ to tune the validation output quality. A test clone pair is reported as true positive if $Pr[\mathbf{y}_t = (1,0)]\geq \gamma$. The default value of $\gamma$ is set to $0.5$ for deciding the clone validation (i.e. classified as true positive code clone if $Pr[\mathbf{y}_t = (1,0)] \geq Pr[\mathbf{y}_t = (0,1)]$. So, on setting this $\gamma$
value towards its upper limit (i.e. 1.0), the proposed method becomes more strict for classifying clones and will return only those clone pairs
having a higher probability of being true positive clones. Thus
most of the returned results are expected to be true code clone
pairs. Similarly one can decrease the value of $\gamma$ to make the proposed method more tolerant for classifying the clones in true positive class. That is this decision
threshold can be useful for the users to tune the result quality
as per the requirements.

% following para should go the cross validation part
%=======================================\\

%following will have to remove possibly
%======\\
%The extracted features of the manually validated
%code clones are fed into the supervised machine learning algorithms for 
%training the model in Step 6. To study the comparative classification performances we applied 13 different machine learning algorithm. The algorithms were tested using 10-fold cross validation method. On an average we found promising classification accuracy
%for some of those algorithms. The detail comparison study of the algorithms has been discussed in \ref{subsec:mult_ml_perform}. 

%=====================================

\subsection{Improving Machine Learning Model with Supervised User Feedback}\label{subsec:improving_ml_models_w_feedback}
The classified code clone pairs from the trained machine learning model are sent back to the corresponding code clone detection tools. User feedback on those test code clone pairs can be collected and stored in a database by the proposed method. The feedback code clone pairs can be optionally used along with the existing training set to further improve the code clone classification. This cycle of supervised learning adds the possibility of improving the classification accuracy over time and experience.\\
\indent The pre-trained model can be used in \textit{transfer learning} \cite{pan2009survey,torrey2010transfer} towards developing custom validation models for given use-cases with the manual validation feedback from the cycle of supervised learning. The goal of the transfer learning is to improve or customize the target task leveraging the knowledge from source task \cite{torrey2010transfer}. Hence, our pre-trained Artificial Neural Network model from large set of training dataset can be used to improve or customize the validation model with comparatively minimal effort.

% We have discussed about the architecture and the prototype implementation in detail in Section \ref{subsec:cloud_arch}.
%\subsection{Comparative Study on Machine Learning Algorithms for Clone Classification} \label{subsec:alg_comp}

\subsection{Cloud Architecture for Clone Classification} \label{subsec:cloud_arch}
\begin{figure}[!htbp]
\centerline{\includegraphics[scale=0.6]{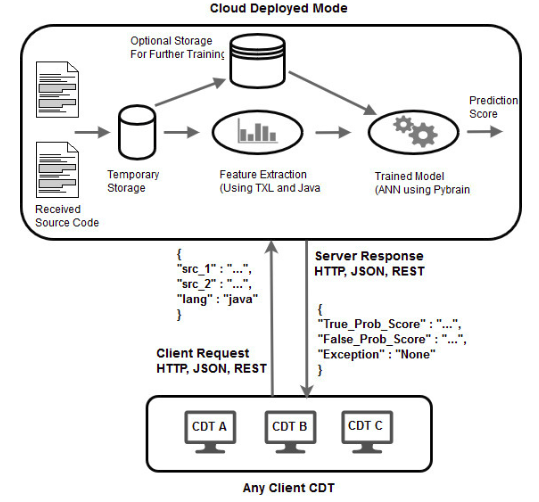}}
\caption{Cloud Model for Compatibility with Existing Code Clone Detection Tools}
\label{fig_proposed_workflow_cloud}
\end{figure}
In addition to using the trained model locally for code clone classification, we also extend the proposed method with a cloud-based architecture for several additional advantages. Machine learning model generalization and performance significantly depends on the quality and quantity of the training data set. With relatively more data, the machine learning models are expected to perform better. As the proposed method can work on top of any clone detection tools generated results, to increase the usability we propose a cloud-based architecture. Figure \ref{fig_proposed_workflow_cloud}, gives an overview of the architecture. Reported code clones from a target code clone detection tool are sent to the server for validation using an HTTP request. The request mainly contains the targeted code clone pairs that need to be validated. The request can optionally contain some additional information to be used by the proposed method. For example, the classification models to use, possible configuration for the classification model and language of the code clone source code. The communication with the server is done using JavaScript Object Notation \cite{crockford2006application} (in key-value pairs as an example shown in Figure \ref{fig_proposed_workflow_cloud}). On receiving the subject clones to validate, the required features are extracted to build the feature vector $\mathbf{x}_t$, which is then used by the trained machine learning model to get the probability score $f(\mathbf{x}_t)$. The corresponding scores are then sent back to the clone detection tool for displaying validated result on the user end.\footnote{We implemented a prototype of the proposed system. All the source codes and required resources will be available after the blind review phase.} Some of the advantages of the cloud deployment of the model can be discussed as the following:
\subsubsection{Compatibility with Existing Clone Detection Tools}
All the platforms and implementation dependencies can be abstracted from the cloud implementation. For example, existing different clone detection tools are developed for different platforms specifically or with different programming languages. Providing a common way (for example the proposed method uses JavaScript Object Notation) that is understandable by all the tools irrespective of their implementation varieties, and can thus improve the usability of the proposed method to a great extent.\\
\indent Listing 6 shows a sample REST API request for automatic clone validation. As it follows standard REST API request format, any of the existing clone detection tool can make such requests for automatic clone validation irrespective of the underlying architecture or implementation languages of the corresponding tools. Getting the validation requests, the proposed validation tool makes all pre-processing of the clone fragments, uses the pre-trained cloud deployed model for making  prediction and finally sends the validation response to the corresponding clone detection tools (i.e., as presented in Listing 7). The architecture thus ensures compatibility with any clone detection tools with minimal efforts.    

\begin{lstlisting}[caption=Sample REST API request for clone validation]

{
    "lang": "Java",
    "sourceCode_1" : "<code clone fragment 1>",
    "sourceCode_2" : "<code clone fragment 2>"
}

\end{lstlisting}

\begin{lstlisting}[caption=Sample REST API response with automatic clone validation]

{
    "output": {
        "prob_false_clone_pair" : 0.1
        "prob_true_clone_pair" : 0.9
    },
    "log_msg" : "Preprocessing clones, Normalizing Codes,...",
    "error_msg" : None
}

\end{lstlisting}

\subsubsection{Improvement in the Training Phase}
Getting enough training data or time by an individual user for the machine learning algorithm can often be challenging. On the other hand, for the cloud based model deployment the user can take advantage of the trained model. Triggering a new model learning on train data set is also simpler and involves a single request to the cloud. Additionally, the user can also choose among different trained models (i.e. Artificial Neural Networks or Decision Tree) for better convergence with their decision. Besides, the cloud-based architecture also adds the possibility of managing a common knowledge base of validation for a user group working on a specific project. For project-specific clone analysis, the project team often targets particular code clones of interest as per the task at hand. In these cases, the cloud-based architecture can be useful for managing the training dataset for common validation patterns.\\
\indent Cloud-based model also opens up the possibilities for higher processing power with cluster or distributed computing for future works. Thus the higher processing advantages for big data of clone validation is possible with even from relatively low processing power end devices for analysis.

\section{Studying Data Distribution for the Clone Classification}\label{sec:data_distribution}
%The proposed approach uses machine learning algorithm that
%over time learns to predict the programmer’s validation
%behaviour and applies that to automatically validate the reported
%list of possible code clone pairs from any code clone detection
%tools. 
In the above Section \ref{sec:proposed_method}, we provided a high-level workflow for the usage of machine for code clone validation. In this section, we present our data distribution study (i.e., in terms of true positive and false positive code clones) for the feature selection of the machine leaning model. We have divided the discussion of this section on dataset
description in three parts. In Section \ref{data_source} we discuss 
the data sources that we used, next Section \ref{data_high_level_details} contains
the discussion on high-level details of the dataset that was
used to train and test the system and finally in Section \ref{data_distribution} we 
discuss about some statistical summaries and underlying
distribution in terms of different extracted features.

\begin{figure}[!htbp]
\centerline{\includegraphics[scale=.7]{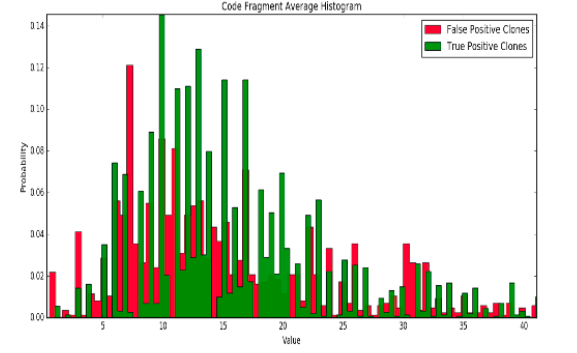}}
\caption{Histogram of Code Fragment Average}
\label{fig_hist_code_frag_avg}
\end{figure}

%need to find a better place for the following information
%====================================================
%\indent To give more generality to the machine learning model we used five different clone detection tools (NiCad \cite{cordy2011nicad}, Deckard \cite{jiang2007deckard}, iClones \cite{gode2009incremental}, CCFinderX \cite{kamiya2002ccfinder} and SourcererCC \cite{sajnani2016sourcerercc}) for detecting possible software clones from the open source projects as shown in Step 2 (Figure \ref{fig_proposed_workflow}). The code clone pairs were then manually validated for the training phase of the proposed method. As some recent research shows that the clone validation decision in some scenario depends on users’ perspective \cite{keivanloo2015threshold}, that is given a possible code clone pair to validate some judges might decide it to be a true positive clone pairs where others might say the opposite (especially in case of Type 3 and Type 4 clones). So to minimize the biasness in the proposed method the whole set of code pairs were split into 5 parts to be validated by 5 different programmers independently (Step 4).
%=====================================================

\subsection{Data Source} \label{data_source}
As the machine learning algorithm tries to recognize any
available underlying pattern in the given dataset, it is 
important how we choose the dataset and which features we
extract out of it for training and testing of the system. For
example, selecting a smaller or undiversified dataset can make
the algorithm biased, resulting in the failure to generalize all the other 
types of clones. So to get generalization in validating different
types of code clones by the system we have chosen to use
a relatively bigger and diverse dataset of open source projects. Besides, we also considered
clones reported by different existing clone detection tools
from those multiple open source projects.\\
\indent For training, we used clones from IJaDataset 2.0 \cite{ijadataset}, - large inter-project dataset of open-source Java systems. To test the
generality of the proposed method, 5 different publicly available
and state-of-the-art tools namely NiCad \cite{roy2008nicad}, Deckard \cite{jiang2007deckard},
iClones \cite{gode2009incremental}, CCFinderX \cite{kamiya2002ccfinder} and SourcererCC \cite{sajnani2016sourcerercc} were used to
detect clones separately out of the benchmark. While we could use a different set of clone detection tools, our target for the generality test suffices by the selected tools. The selected clone detection tools are widely used and researched in recent time, they provide a common ground for the experimental evaluation. In addition to that, as the proposed method works directly on the clone fragments (i.e., for feature extraction), it should be compatible with any clone detection tools, as all of the tools return some meta-data or information about the clone fragments from the original software system. \\
\indent Randomly, 400
clone pairs were then selected and manually validated from each
of the five clone detection tools separately. We have chosen to
work on these dataset because a good number of recent research
works on code clones has been carried out on these open source
projects and thus we can have a common ground for evaluating
the proposed approach.
\subsection{High-Level Details of the Data Set} \label{data_high_level_details}
Reports obtained from any of the existing clone detection tools
on possible code clone pairs are given as input to the proposed
method for validation purposes. Several code clone detection
tools were run on the used data source to find the corresponding
reports for the possible code clone pairs. The code clone pairs were then manually validated for the
training phase of the proposed method. As some recent research
shows that the clone validation decision in some scenario
depends on user's perspective \cite{keivanloo2015threshold}, that is given a possible code
clone pair to validate some judges might decide it to be a true
positive clone pairs where others might say the opposite
(especially in case of Type 3 and Type 4 clones). So to consider this
generalization to the proposed method the whole set of code
pairs were split into five parts to be validated by five different
graduate research students from computer science background. This manual validation decision
along with the corresponding possible code clone pairs are given
as input to the proposed method for the training phase.

\begin{figure}[!htbp]
\centerline{\includegraphics[scale=.7]{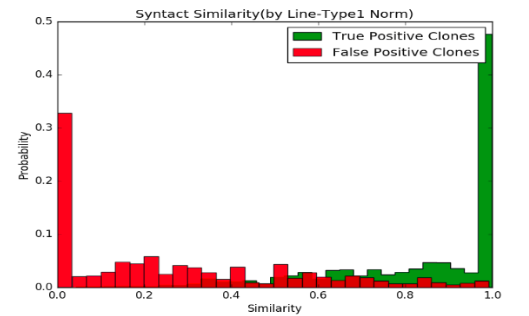}}
\caption{Histogram of Syntactical Similarity by Line (Type 1 Norm.)}
\label{fig_hist_syntact_line_type1}
\end{figure}

%\subsection{Feature Extraction Methodologies}

\subsection{Analyzing Data Distribution for the Clone Classification Problem} \label{data_distribution}
Out of those manually validated clones we extracted
different features that are used to train the machine learning model. In this section, we 
discuss different distribution and statistical studies and
behaviors of some of the extracted features.\\
\indent For every code clone pair detected by clone detection tools,
we found the similar code fragments for a clone pair. These are the similar code fragments for which the tools decided could be a code clone pair.
%==== fragment size diff ==== add some intro....\\
We analyzed
this feature of the code clone pairs for both true positive and
false positive manually validated clones in an attempt to find its 
contribution score for clone classification. Figure \ref{fig_hist_code_frag_avg}, shows the distribution
of the average code clone fragment feature ($(\alpha+\beta)/2$, as discussed in Section \ref{subsec:feature_extraction}) for the true positive
and false positive clone classes. From the figure, we can notice
that the average code fragment size shows much randomness,
both for true positive and false positive clones. The distribution
of this feature almost overlaps on one another for the two
classes: true positive and false positive code clones. This
overlapping pattern suggests that this feature provides very minimal information about the two classes and thus yields a very low possible
contribution score for training the machine learning algorithm for validation.\\
\begin{figure}[!htbp]
\centerline{\includegraphics[scale=.6]{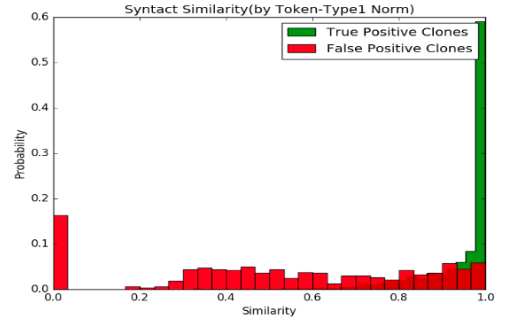}}
\caption{Histogram of Syntactical Similarity by Token (Type 1 Norm.)}
\label{fig_hist_syntact_token_type1}
\end{figure}
\indent Besides, for extracting some other features, we 
normalized the code clone pairs by 3 levels,
namely: Type 1, Type 2 and Type 3. Then for each
level of normalizations, the syntactical similarity was measured by
lines and by tokens for the clone pairs resulting in 6 different possible
features (Section \ref{subsec:feature_extraction}). To view any underlying distribution of the features
their normalized histogram were plotted both for true positive
and false positive clones. Figure \ref{fig_hist_syntact_line_type1}, shows one such plottings that
is based on the similarity measured by lines after Type 1 code
normalization. From the figure, it is noticeable that the
distribution of the feature is comparatively better than the
average code fragment line feature in terms of validation.
Though the distribution for true positive and false positive
clones are not completely linearly separable with this feature but
still the two classes are somewhat distinguishable. The distribution indicates a better possible 
 contribution score for validation prediction than the average clone fragment sizes. Figure \ref{fig_hist_syntact_token_type1}, also shows somewhat similar results in the case of
Syntactic Similarity measured by tokens after Type 1
Normalization of the source codes.\\
\begin{figure}[b]
\centerline{\includegraphics[scale=.65]{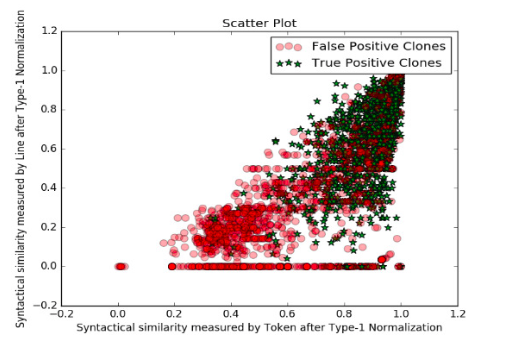}}
\caption{Syntactical Sim. by Line vs Token (Type 1 Norm.)}
\label{fig_syntact_sim_line_vs_token}
\end{figure}
\indent We also carried out several studies to find out any underlying
relationships between different features for possible clustering of the two clone  classes. For example, we tried to figure out if there is any underlying
relationship available for different types of similarity measures
that can give any potential information about the clustering
of the two clone classes. We plotted our several study results for
visualization in an attempt to notice any distinguishable
separation or clusters. For example, Figure \ref{fig_syntact_sim_line_vs_token}, is one such study results that shows
the scatter plot on syntactical similarity measured by line versus
tokens after Type 1 Normalization of the code clone pairs. However, these
analyses did not show any distinguishable cluster information for the two
classes.\\
\indent As machine learning algorithms try to recognize any
underlying pattern available on the working dataset, the detailed 
analysis on the dataset and possible features are necessary for
selecting the right machine learning algorithm and features. This distribution analysis on different possible features for
code clones provides information about their importance and
contribution for clone validation. This analysis provides a clearer
view of the data distribution and thus helps to pick the
appropriate machine learning algorithm and corresponding
features for the algorithm. From several analyses on the data
distribution, we tried to find out the features that have
comparatively more distinguishable distribution and provides
more contribution for the two classes – true positive and false
positive clones. Table \ref{table:selected_features}, shows a feature set ranked on possible contribution score based on our analysis study. The corresponding distribution mean differences, $\Delta\mu$ for the two
classes also somewhat indicates the separability for the classification.\\
\indent The detailed feature study, in terms of class distribution prior to applying any machine learning algorithm is very important, since using any noisy feature (for the specific classification problem) may affect the classification performance and reduces the generality of the classification. The distribution study,  also contributes to the research area for further improvement in feature extraction and selection of appropriate classification algorithms. From our study, we built the feature vector as listed in Table \ref{table:selected_features}. The other features were not used for the clone classification due to their low contribution scores or noisy behaviors for the classification as discussed above. 
\begin{table}[!htbp]
%% increase table row spacing, adjust to taste
\renewcommand{\arraystretch}{1.3}
% if using array.sty, it might be a good idea to tweak the value of
% \extrarowheight as needed to properly center the text within the cells
\caption{Selected Features Based on Distribution Analysis}
\label{table:selected_features}
\centering
%% Some packages, such as MDW tools, offer better commands for making tables
%% than the plain LaTeX2e tabular which is used here.
\resizebox{\textwidth}{!}{\begin{tabular}{l|c|l}
\hline
\textbf{Feature} & \textbf{$\Delta\mu$} & \textbf{Feature Summary (as discussed in details in Section \ref{subsec:feature_extraction}})	\\
\hline
\hline
Line Sim. (Type-1 Norm.) &	0.3998 & Syntactical similarity measured by line after Type-1 Normalization \\
\hline
Line Sim. (Type-2 Norm.) &	0.3701 & Syntactical similarity measured by line after Type-2 Normalization \\
\hline
Line Sim. (Type-3 Norm.) &	0.3602 & Syntactical similarity measured by line after Type-3 Normalization \\
\hline
Token Sim. (Type-2 Norm.) &	0.3447 & Syntactical similarity measured by Token after Type-2 Normalization \\
\hline
Token Sim. (Type-1 Norm.) &	0.3105 & Syntactical similarity measured by Token after Type-1 Normalization \\
\hline
Token Sim. (Type-3 Norm.) &	0.2537 & Syntactical similarity measured by Token after Type-3 Normalization \\
\hline
Function Intersected &	0.2364 & Total Number of functions intersected by the code fragments \\
\hline
Unmatched Braces &	0.2078	& Total number of unmatched braces across both code fragment \\
\hline
\end{tabular}}
\end{table}

\begin{figure}[!htbp]
\centerline{\includegraphics[scale=0.5]{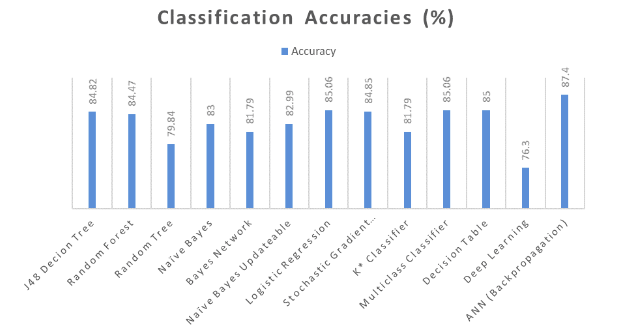}}
\caption{Accuracy Comparison of the Methods across Different Software Systems.}
\label{fig_alg_acc}
\end{figure}

\section{Performance Analysis of Different Machine Learning Models: A Comparative Study}\label{subsec:mult_ml_perform}
In the last Section \ref{sec:proposed_method} and Section \ref{sec:data_distribution} we presented the high-level workflow involving machine learning and data distribution analysis for feature selection by the classification models respectively. In this section, we study the performance of different machine learning models in our proposed high-level workflow with the studied feature set.
%\indent The data distribution study (Section \ref{data_distribution}) exhibits that the two classes are not linearly separable or do not contain any distinguishable clone clusters. The study shows that the clone classes are not straightforward or linear for separation. For example by the usage of single or multiple thresholds \cite{keivanloo2015threshold} in terms of the features by code clone detection tools \cite{kamiya2002ccfinder}, \cite{baker1993program}. On the other hand, somewhat recognizable distribution for the two classes for some of the features as discussed in Section \ref{data_distribution}, indicates the possibility of having complex underlying patterns or rules available in the data set that can contribute to the classification problem of clone validation. So our intuition was that, if we can improve the tool generated report  on  code  clone  by  exploiting  those complex patterns and the  user  preferences by mapping the extracted feature set to  the  corresponding  subjective  clone  validation problem. In  an attempt  for  such  improvement  on  clone  reporting  we  applied and  studied  different  machine  learning  algorithms  on  the open  source  software  projects (i.e. the training dataset as discussed in Section \ref{data_source}).
\subsection{Bayes Classifiers}
From the extracted code clone feature vector $\mathbf{x}=(x_1, x_2, \cdots , x_n)$, we experimented with the Naive Bayes Classifier - a conditional probability model, for classification of the clonesets into two clone classes - $C_T$ and $C_F$, representing true and false positive validated clone classes respectively. For the extracted $n$ clone features of a reported clone pair, the classifier assigns conditional probabilities for the two classes - $Pr(C_T|x_1, x_2, \cdots , x_n)$  and $Pr(C_F|x_1, x_2, \cdots , x_n)$ using the Bayes' Theorem as Eq. \ref{eq:bayesTheorem}: 
\begin{equation}\label{eq:bayesTheorem}
Pr(C_k|\mathbf{x}) = \frac{Pr(C_k)Pr(\mathbf{x}|C_k)}{Pr(\mathbf{x})}
\end{equation}
where, $k=\{T,F\}$, $Pr(C_k)$ is the prior probability of the clone class $C_k$, $Pr(\mathbf{x}|C_k)$ is the likelihood of the clone pair with feature vector $\mathbf{x}$ to be in the clone class $C_k$ and $Pr(\mathbf{x})$ is the evidence of the feature vector $\mathbf{x}$. The evidence can be ignored as it is independent of the clone class $C_k$. Under the assumption that the clone feature $x_i$ is independent of any other feature $x_j$ for $i\neq j$, Naive Bayes Classifier then assigns the class probabilities for a given test feature vector $\mathbf{x_t}$ as Eq.  \ref{eq:bayesTheorem_classification}:
\begin{equation}\label{eq:bayesTheorem_classification}
Pr(C_k|\mathbf{x_t})= Pr(C_k)\prod_{i=1}^n Pr(x_i|C_k) 
\end{equation}
\begin{figure}[tb]
\centerline{\includegraphics[scale=0.6]{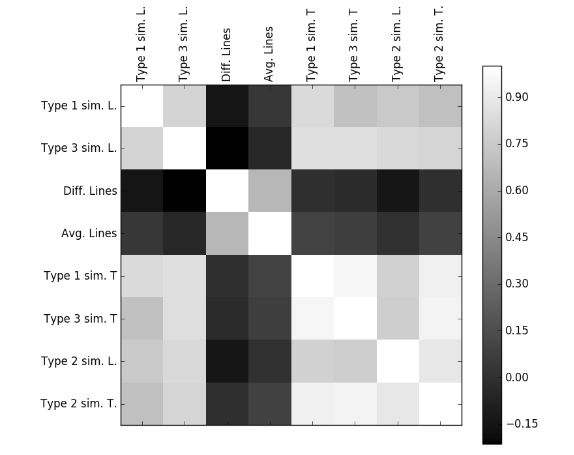}}
\caption{Correlation among a feature Subset.}
\label{fig_feature_corr}
\end{figure}
We used kernel density estimation \cite{john1995estimating} for the likelihood calculation as most of the selected feature values are continuous. With described configurations the classifier showed an accuracy of 83\%, with 0.831 and 0.830 of precision and recall respectively (More result quality analysis reports in Table \ref{table:res_quality}).\\ 
\indent However, it is a strong assumption by the classifier to consider independence among the extracted features for a given class $C$, for the clone classification problem. Because, by the definition of the code clones \cite{roy2007survey}, it is usual that part of a Type 2 clone can contain Type 1 clone. Similarly, Type 3 code clone can also contain fractions of Type 2 or Type 1 clones. So, by induction it is expected that the extracted similarity features (Section \ref{data_distribution}) for a given clone class have some sort of correlation among them. For example, Figure \ref{fig_feature_corr}, shows the correlation among some of the extracted features. As it is noticeable from the figure, the clones structural features such as average line or line differences shows relatively lower correlation with other features. However, the extracted similarity based features among clone pairs after different levels of normalization shows significantly higher correlation among them. From these findings, we experimented with the Bayesian Network Classifier \cite{friedman1997bayesian} - that considers and learns the possible dependency relations among the features. Unsupervised learning was used (via Minimum Description Length (MDL) \cite{lam1994learning} scoring method) to build the dependency network.\\ 
\indent Though Bayesian Network tries to mitigate the strong assumption made by the Naive Bayes, we found that the two classifiers perform relatively the same for the clone classification problem with the used features. In fact, in some cases, naive Bayes outperformed the Bayesian Classifier (as shown in Figure \ref{fig_alg_acc}). This behaviour is not totally unexpected though, as Friedman et. al. \cite{friedman1997bayesian} showed a detailed study on this. Error while learning the dependency network from the training set was presented as possible reasoning for such behaviour.

\begin{table*}[t]
%% increase table row spacing, adjust to taste
\renewcommand{\arraystretch}{1.3}
% if using array.sty, it might be a good idea to tweak the value of
% \extrarowheight as needed to properly center the text within the cells
\caption{Classification Result Quality for Different Machine Learning Algorithms}
\label{table:res_quality}
\centering
%% Some packages, such as MDW tools, offer better commands for making tables
%% than the plain LaTeX2e tabular which is used here.
\resizebox{\textwidth}{!}{\begin{tabular}{l | c | c | c | c | c | c}
\hline
\textbf{Classifiers} & \textbf{TP Rate} &	\textbf{FP Rate} & \textbf{Precision}	& \textbf{Recall} & \textbf{F-Measure} & \textbf{ROC-area} \\
\hline
\hline
J48 Decision Tree &	0.848 &	0.291 &	0.849 &	0.848 &	0.840 &	0.803 \\
\hline
Random Forest &	0.845 &	0.254 &	0.841 &	0.845 &	0.841 &	0.892 \\
\hline
Random Tree	& 0.789	& 0.275	& 0.799	& 0.798	& 0.799	& 0.793 \\
\hline
Naive Bayes Classifier	& 0.830	& 0.332	& 0.831	& 0.830	& 0.818	& 0.828 \\
\hline
Bayes Network	& 0.818	& 0.266	& 0.815	& 0.818	& 0.816	& 0.830 \\
\hline
Naive Bayes Updateable	& 0.830	& 0.332	& 0.831	& 0.830	& 0.818	& 0.828 \\
\hline
Logistic Regression	& 0.851	& 0.292	& 0.852	& 0.851	& 0.842	& 0.845 \\
\hline
Stochastic Gradient Descent	& 0.849	& 0.308	& 0.854	& 0.849	& 0.838	& 0.770 \\
\hline
K* Classifier	& 0.818	& 0.287	& 0.813	& 0.818	& 0.814	& 0.848 \\
\hline
Multiclass Classifier	& 0.851	& 0.292	& 0.852	& 0.851	& 0.842	& 0.845 \\
\hline
Decision Table	& 0.850	& 0.292	& 0.852	& 0.850	& 0.841	& 0.845 \\
\hline
\end{tabular}}
\end{table*}

\subsection{Decision Tree Classifiers}
For predicting the target variables, these classifiers build a decision tree from the input variables of the used feature vector $\mathbf{x}$. The internal nodes of the tree correspond to different input variables, values of the corresponding input variables define the edges connecting nodes and each leaf denotes different target variable for the classification. At each step of learning the model, an input variable is selected as a node, such that it best splits the remaining training data set. There are several variations of the classification based on this recursive training set split scoring and corresponding node selection. We used a number of them for testing the performance for the clone classification problem. The Pruned C4.5 decision tree \cite{Quinlan1993} showed an accuracy of 84\%. The obtained precision and recall are 0.849 and 0.848 respectively. Random Tree is also another variation of the classification group that considers $K$ random input variables at steps for generating the decision tree. The obtained accuracy was 79\%. The precision and recall values were also relatively lower than C4.5 decision tree. We also experimented with Random Forest classifier \cite{Breiman2001}, that considers multiple tree decisions for building the model. We got approximately similar accuracy with this classification algorithm as C4.5 decision tree. However, precision and recall values show relatively lower values than C4.5 decision tree.

\subsection{Backpropagation Neural Network}
\begin{figure}[tb]
\centerline{\includegraphics[scale=.7]{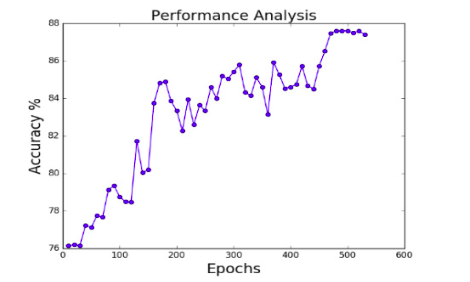}}
\caption{Avg. Accuracy for 10-fold cross-validation (as the algorithm converges vs epochs).}
\label{fig_performance_analysis}
\end{figure}

From the training dataset $D$, for $\mathbf{x}_i\in \mathbb{R}^n$ in the input layer, $\mathbf{y}_i\in \mathbb{R}^l$ in the output layer and one hidden layer with $k$ nodes, the ANN learns the following function:
\begin{equation}
f(\mathbf{x}) = \sigma(W^\intercal_{ho}\cdot\sigma(W^\intercal_{ih}\cdot \mathbf{x} + \theta_h) + \theta_0)
\end{equation}
where, $W_{ih} \in \mathbb{R}^{n\times k}$ and $W_{ho} \in \mathbb{R}^{k\times l}$ denotes the connection weights from the input layer to the hidden layer and hidden to output layer respectively. $\theta$ and $\sigma$ denote the layer bias and neuron activation function respectively. We used softmax activation function for the output layer. The learned function $f(\mathbf{x})$, is then used to predict the clone validation for the new test feature vector $\mathbf{x}_t$ from the corresponding probability values:
\begin{equation}
\hat{\mathbf{y}_t} = f(\mathbf{x}_t) = (Pr[\mathbf{y}_t = (1,0)], Pr[\mathbf{y}_t = (0, 1)])
\end{equation}
where, $Pr[\mathbf{y}_t = (1,0)]$ denotes the probability of the test code clone with feature $\mathbf{x}_t$ to be true positive. So, for a preset user preference value $\gamma[0,1]$, the proposed approach decides the test code clone as True Positive if $\lambda\geq\gamma$ and False Positive otherwise (as discussed in Section \ref{subsec:prediction_decision_config}).\\
\indent For the training phase, the Neural Network was run with different values of $k$ (to investigate the optimal network configuration), for a number of
epochs until it converges with a maximum limit of 1000 epochs.
The model was trained and tested using 10-fold cross
validation. Figure \ref{fig_performance_analysis}, shows the accuracy of the method as it
converges versus the epochs (averaged for each of the 10-fold
validation). The Neural Network converged within a range of
500 to 600 epochs for $k=107$, giving an accuracy of 87.4\%.

\begin{figure}[!htbp]
\centerline{\includegraphics[scale=.7]{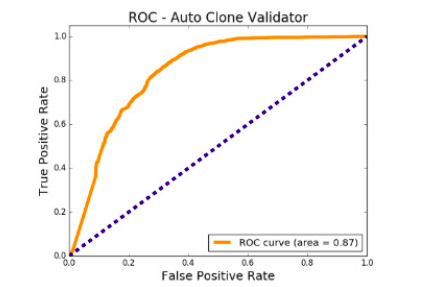}}
\caption{ROC - Curve for validation by the method}
\label{fig_roc}
\end{figure}

\begin{figure}[!htbp]
\centerline{\includegraphics[scale=.7]{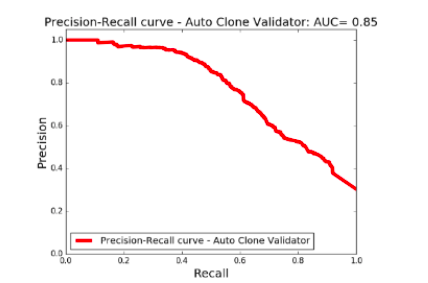}}
\caption{PR - Curve for validation by the method}
\label{fig_pr}
\end{figure}

%===========\\
To analyze the output quality across different values of $\gamma$ we
plotted the ROC curve, which is shown in Figure \ref{fig_roc}. The
calculated Area Under the Curve (AUC) for the ROC curve is
found to be 0.87. From the ROC curve, the proposed method can recommend the $\Theta$ value to the users by default that gives the best
result in terms of the ratio of the true positive and false positive
ratio in the output results while training. Besides Figure \ref{fig_pr}, shows the Precision-Recall curve
for the proposed method for varying $\gamma$[0,1] values. In case of
Precision-Recall curve, the AUC found to be 0.85.\\

\subsection{Deep Learning}
In addition to the Backpropagation Neural Network, we also explored extending the model to deep learning for training the prediction function $f(\mathbf{x})$, for mapping $\mathbb{R}^n$ to $\mathbb{R}^l$, where $n$ and $l$ are dimensions of feature vector and class labels respectively. As per Equation 7, we applied the Rectified Linear Unit (ReLU), $\sigma(z) = \max(0, z)$ and Sigmoid function, $\sigma(z) = \frac{1}{1 + \exp(-z)}$ as the activation functions for the hidden and output layers respectively. Deep Neural Networks with a large number of layers and nodes often face an overfitting problem \cite{srivastava2014dropout}. In these cases, some specific set of neurons adapts too much in the decision while ignoring a large set of other neurons, and thus failing to generalize the learned classification model. Srivastava et. al. \cite{srivastava2014dropout} proposed a dropout method for preventing neural networks from  such an overfitting problem. This method drops out random nuerons along with their corresponding connections forming thinned networks in the training phase preventing too much co-adaption. The network is then approximated from the thinned networks in the testing phase.\\
\indent We used Keras \cite{keras} - a python deep learning library, running on top of TensorFlow \cite{abadi2016tensorflow} - a recent open source project released by Google for deep learning. We investigated the model's clone classification performance with different configurations (e.g. different dropout probability, learning nodes, hidden layers). From our investigation, we got a comparatively better result with the sequential stacking of three layers in addition to the input and output layers. We used dropout probability $p=0.5$ (e.g. a neuron along with it's corresponding connection is dropped out with a probability of $0.5$), giving generality in training the model. With 32 neurons in each of the hidden layers, the obtained accuracy was 76\%.
\\
\\
\indent The data distribution study in Section \ref{data_distribution}, indicated the non-linear function requirement for clone classification. Hence from the findings, we investigated different machine learning models for the classification. The study helps better understanding the classification problem and also can contribute to future research works in this area for building even better models from our insights. We got the best classification result using Backpropagation Neural Networks. To verify the classification performance for different use cases, we have performed several experiments with this machine learning model. The detail study findings have been presented in Section \ref{sec:exp}.

\section{Experiments}\label{sec:exp}
\subsection{Implementation Details}\label{implementation_details}
\begin{figure}[t]
\centerline{\includegraphics[scale=0.4]{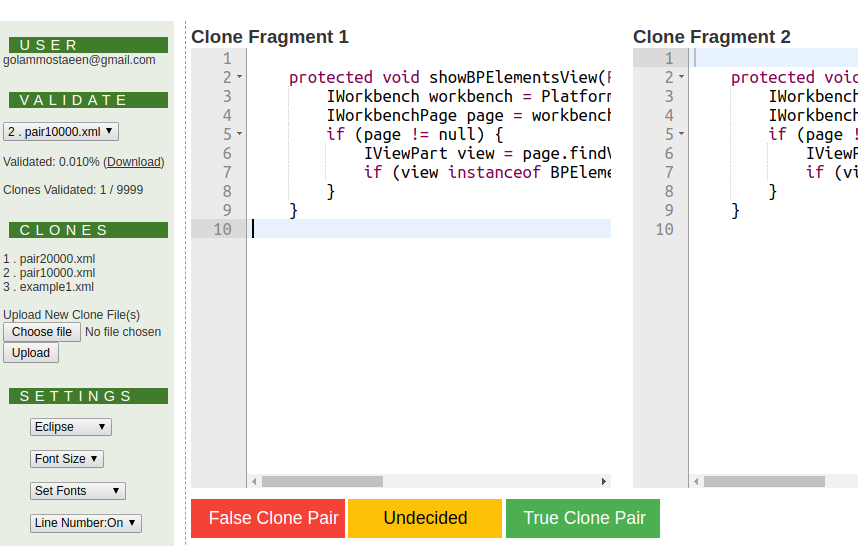}}
\caption{System Prototype: Systems Clone Validation}
\label{fig_proposed_imp_clone_validation}
\end{figure}

\begin{figure}[!htbp]
\centerline{\includegraphics[scale=0.5]{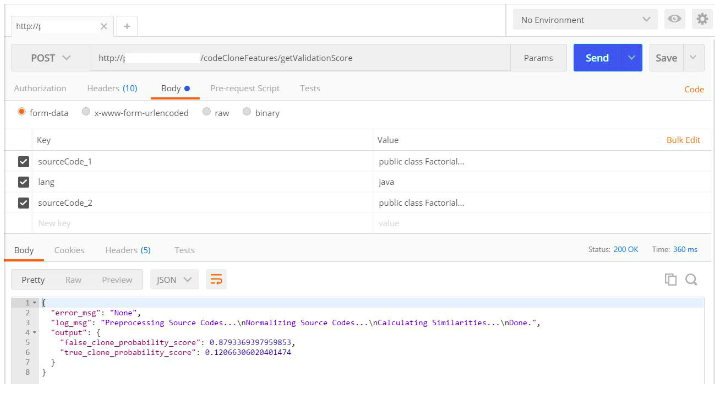}}
\caption{System Prototype: Getting Prediction Response from Cloud}
\label{fig_proposed_prediction_response}
\end{figure}

We implemented a prototype\footnote{https://github.com/pseudoPixels/CloneCognition} of the system for testing the performance of the proposed method in different experimental setups. For collecting the user-specific training data, a cloud-based web application was first developed as shown in Figure \ref{fig_proposed_imp_clone_validation}. We used Python 2.7, as the server-side language. The web application was developed using Flask\cite{flask} - a microframework for python. The system server can be populated by code clones reported by different code clone detection tools for user-specific validation. The system iteratively displays the code clones to the users for manual validation. For a given code clone pair, user decisions (true positive or false positive) are then stored in the server database, mapping against the corresponding user profile. We used CouchDB\cite{couchdb} - a NoSQL database system, that supports easier scaling up and distributed computing for Big Data \cite{moniruzzaman2013nosql}. We selected CouchDB to take advantage of this feature of the database for handling a large amount of code clones in our future works.\\
\indent The collected manual validated code clone pairs are then used for building the classification model as described in the proposed method. We used TXL \cite{cordy1991txl} for pre-processing the code clone pairs for extracting several features (as described in Section \ref{data_distribution}) for training the model. Following the pre-processing, the clone features were then extracted using the Java programming language. The feature vectors from the training dataset were then used for learning the classification model. We experimented with several machine learning algorithms for the clone classification. The details of the findings and comparison study among the classifiers have been presented in Section \ref{subsec:mult_ml_perform}. The trained model is then used for user-specific validation of new code clone pairs. The reported code clone pairs for validation can be sent to the cloud, where the trained model predicts and returns the validation score to the corresponding user end. We also implemented a prototype for receiving the test code clone pairs and sending the validation score from trained model to the user end. As shown in Figure \ref{fig_proposed_prediction_response}, the server is requested with code clone pairs for validation, server then uses the trained machine learning model for clone classification and then sends back the validation score to the user in JSON format. The server requests and responses are done using the RESTful API. The validation score then can be used in the corresponding code clone detection tools for classification or comprehension of the clones as per the user configuration. In addition to this cloud deployment, the modular trained model can also be used locally for prediction by embedding with particular clone detection tools.\footnote{The prototype system implementation and source codes will be made available for usage and research purpose after the blind review process.}

\begin{figure}[t]
\centerline{\includegraphics[scale=0.4]{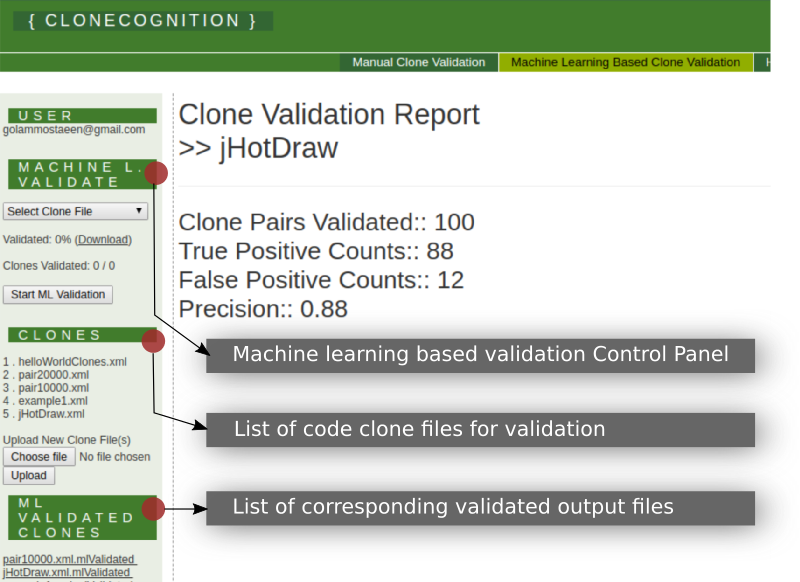}}
\caption{System Prototype: Validated Clone Report.}
\label{fig:cloneReport}
\end{figure}

%write above about the txl operations used

\begin{table}[!htbp]
%% increase table row spacing, adjust to taste
\renewcommand{\arraystretch}{1}
% if using array.sty, it might be a good idea to tweak the value of
% \extrarowheight as needed to properly center the text within the cells
\caption{Information of the open source projects in the experimentation}
\label{table:system_info}
\centering
%% Some packages, such as MDW tools, offer better commands for making tables
%% than the plain LaTeX2e tabular which is used here.
\begin{threeparttable}
\begin{tabular}{l|c|c|c|c}
\hline
%\multirow{2}{*}{\textbf{Software System}} & \multirow{2}{*}{\textbf{LoC clones}}\tnote{1} & \multirow{2}{*}{\textbf{Avg. Lines}}\tnote{2} & \multirow{2}{*}{\textbf{Avg. Tokens}}\tnote{2} & \multicolumn{2}{c|}{\textbf{FICA}}  & \multicolumn{2}{c|}{\textbf{FICA Iterative}} & \multicolumn{2}{c}{\textbf{Proposed Method}} \\
%\cline{5-10}
% & & & & \textbf{Precision} & \textbf{Recall}  & \textbf{Precision} & \textbf{Recall}  & \textbf{Precision} & \textbf{Recall} \\
%\hline
%\hline
\textbf{Software System} & \textbf{LoC}\tnote{1} & \textbf{Clone Pairs} & \textbf{Average Lines}\tnote{2} & \textbf{Average Tokens}\tnote{2} \\
\hline
\hline
Luaj & \textbf{36155} & 1181 & 15 & 79 \\
\hline
Ucdetector & 4388 & 183 & 11 & 67 \\
\hline
Autocover Tool & 3989 & 150
 & 13 & 50 \\
\hline
Upm-swing & \textbf{13243} & 561 & 11 & 73 \\
\hline
ipscan & 7082 & 323 & 10 & 58 \\
\hline
JavaGB & 24211 & 1246 & 9 & 58 \\
\hline
JavaOcr & 7699 & 208
& 18 & \textbf{90} \\
\hline
JavaFileManager & \textbf{25898} & 1017
& 12 & 68 \\
\hline
jMemorize & 13109 & 598 & 10 & 44 \\
\hline
FileBot & 18369 & 834 & 11 & 59 \\
\hline
JAIMBot & 14096 & 583 & 12 & \textbf{83} \\
\hline
JLipSync & 3671 & 64 & 28 & 158 \\
\hline

\end{tabular}

\begin{tablenotes}
    \item[1] Some of results are combination of detected clones from multiple clone detection tools (as listed in Table \ref{table:clone_detection_tools})
    \item[2] Average per code clone fragment
\end{tablenotes}
\end{threeparttable}
\end{table}

\subsection{Experimental Setup}\label{subsec:exp_setup}
Automatic clone validation can contribute to clone analysis
across different scenarios and requirements 
starting from smaller to large scale software system. For this
reason, we were interested in evaluating the system across
different environmental set ups: with several clone detection
tools, users, and software systems. Table \ref{table:system_info} lists a set of open-source projects that we used in evaluating the proposed system. For testing the model generality, we also used multiple clone detection tools on those software systems for detecting code clones. Table \ref{table:clone_detection_tools} shows the clone detection tools along with their used configuration for the study.

\begin{table}[!htbp]
%% increase table row spacing, adjust to taste
\renewcommand{\arraystretch}{1.3}
% if using array.sty, it might be a good idea to tweak the value of
% \extrarowheight as needed to properly center the text within the cells
\caption{Used Clone Detection Tools for the Study}
\label{table:clone_detection_tools}
\centering
%% Some packages, such as MDW tools, offer better commands for making tables
%% than the plain LaTeX2e tabular which is used here.
\resizebox{\textwidth}{!}{\begin{tabular}{l|c|l}
\hline
\textbf{CDT} & \textbf{Ver.} & \textbf{Tool Configuration}	\\
\hline
\hline
iClones \cite{gode2009incremental} &	0.2 & mintokens = 50, minblock = 20 \\
\hline
NiCad \cite{roy2008nicad} &	4.0 & blocks, 30\%, 6-2500 lines, blind-renaming, abstract-literal \\
\hline
SimCad \cite{uddin2013simcad} & 2.2 & generous, 6+ lines, blocks \\
\hline
CloneWorks \cite{svajlenko2017cloneworks} & 0.2 & Type-3 Aggressive, 6 lines, blocks \\
\hline
Simian \cite{simian} & 2.4 & 6 lines, ignore overlapping blocks, balances parentheses \\
\hline
Ctcompare \cite{toomey2012ctcompare} &	3.2 & 50 tokens, 3 replacements \\
\hline
\end{tabular}}
\end{table}

\begin{table}[!htbp]
\caption{Operations used to Create Artificial Code Clones via Mutation Framework \cite{svajlenko2013mutation}}
\begin{center}
\begin{tabular}{c|l}
\hline
\textbf{Clone Types}& \textbf{Modification Operations}\\
\hline
\hline
\multirow{3}{*}{Type-1} & Addition/Removal of white-space \\
\cline{2-2}
 & Changing the code comments\\
\cline{2-2} 
 & Addition/Removal of newlines\\
 \hline
 
 \multirow{3}{*}{Type-2} & Systematic renaming of identifiers \\
\cline{2-2}
 & Arbitrary renaming of identifiers\\
\cline{2-2} 
 & Change in value of literals\\
 \hline
 
 \multirow{3}{*}{Type-3} & Insertion/Deletion within lines \\
\cline{2-2}
 & Insertion/Deletion of lines\\
\cline{2-2} 
 & Modification of whole lines\\
 \hline
\end{tabular}
\label{table:mutation_operations}
\end{center}
\end{table}

\begin{table}[!htbp]
\caption{Result on Artificial Code Clones}
\begin{center}
\begin{tabular}{c|c|c|c}
\hline
\textbf{Accuracy} & \textbf{Precision} & \textbf{Recall} & \textbf{F1-Score}\\
\hline
\hline
90\%  & 0.89 & 0.99 & 0.93\\
\hline
\end{tabular}
\label{table:artificial_clone_result}
\end{center}
\end{table}

\subsection{Evaluation on Artificial Clones}
Evaluation of code clone related tools and techniques can often
be critical as the validation of some of the types of code clones as
true or false positive varies significantly from person to person
\cite{keivanloo2015threshold}, \cite{charpentier2015empirical}. Thus, in order to
get more concrete information about the validation accuracy of
the trained model, we were interested in evaluating the system
with artificially generated clones before testing on real clones
from different software systems. We generated a large number of
true positive clones with all the different kinds of modifications
of the original source codes that possibly generate code clones.
We used the Mutation Framework \cite{svajlenko2013mutation} for creating such a code clone
benchmark. The framework takes a code fragment as input and 
performs mutation operation by random edit operations  on the code
fragments to artificially create a clone pair. We used 9 different mutation edit operations on the source codes as listed in Table \ref{table:mutation_operations}. These operations
create three different types (Type-1, Type-2 and Type-3) of true
positive clones which are mostly simpler, straight forward and
have minimal subjective bias. We used
different original code fragments from BigCloneBench \cite{svajlenko2014towards} to
create 3750 such artificial true positive code clone pairs. Our
target was to test the performance of the proposed method on
validating those artificial true positive clones. So along with
them, we mixed 840 randomly selected false clones from the dataset
as described in Section \ref{data_source}. We then applied the proposed
method on the clones for validation. We got comparatively
better accuracy on these artificially created clones as shown in
Table \ref{table:artificial_clone_result}. The possible reasoning for this is that though the artificially
created clones contain minimal subjective biases, they have one
disadvantage: they are very similar with one another and
comparatively easily distinguishable (as also noticeable from
higher recall value in Table \ref{table:artificial_clone_result}).

\begin{figure}[!htbp]
\centerline{\includegraphics[scale=0.52]{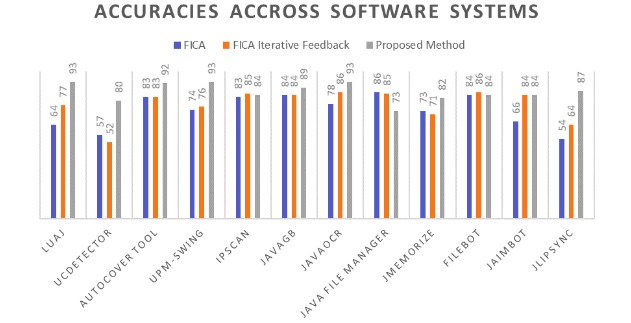}}
\caption{Accuracy Comparison of the Methods Across Different Software Systems.}
\label{fig_sys_acc}
\end{figure}

\subsection{Evaluation on different software systems}
\begin{figure}[!htbp]
\centerline{\includegraphics[scale=.5]{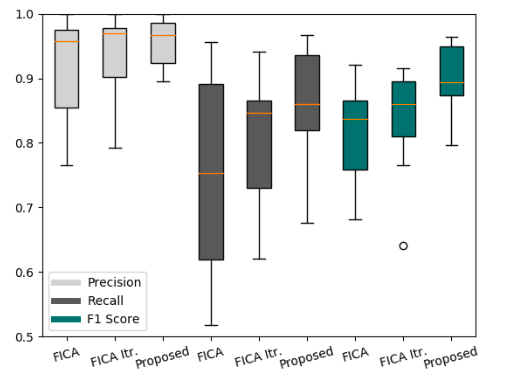}}
\caption{Result Quality Comparison of the Methods}
\label{fig_pr_cmp}
\end{figure}

The proposed method shows a promising result with an accuracy
of 87.4\% via 10-fold cross validation on the data set as discussed
in Section \ref{subsec:mult_ml_perform}. The result also exhibits confidence as the used
dataset is comparatively larger and contains a number of diverse
software projects. However, we were also interested to see how
the proposed method works for different software
projects.
We selected 12 completely different open source projects that
were not used in any of the previous training or testing phases. The used open source software projects have been listed with details in Table \ref{table:system_info}. 
We used different code clone detection tools for detecting the code clones available in those
open source software project. We used 6 different code clone detection tools to test the 
generality of the proposed method. Table \ref{table:clone_detection_tools} lists
the code clone detection tools used along with their corresponding version number and used configurations. The reported
code clones from code clone detection tools were then manually validated by different users. None of the users were
previously involved in building the training dataset as discussed in Section \ref{data_source}. Besides,
to compare the performance of the proposed method with similar existing method - FICA \cite{yang2015classification}, we contacted and
got the source code\footnote{Authors of FICA made the source code available for research purpose at https://github.com/farseerfc/fica} from the corresponding authors.\\
\indent The trained model was used for
predicting the user clone validation for each of the projects. Figure \ref{fig_sys_acc} shows the comparative accuracies for the existing and proposed approaches for different software systems. As noticeable from the graph the proposed approach showed better accuracies for most of the systems. For \lq Java File Manager\rq  , however, unlike the other systems, the proposed approach showed noticeably lower performance. We found the considered clones for the system are mostly Type 1 and Type 2 - which may be a possible reason for such a comparative result for the system.\\
\indent Besides, to test the result quality, system-wide precision and recall were calculated for the approaches. The obtained result has been presented in Table \ref{table:fica_comparison2}. As some of the values have been highlighted in the table, it is noticeable that in most of the cases the precision and recall values get lower in comparison to the proposed approach. The result is also noticeable in the box plot in Figure \ref{fig_pr_cmp}. The box plot illustrates that the mean Precision, Recall or $F_1$-Score for the existing approaches are relatively lower than the proposed. Besides, the plot also depicts a higher variation in the result qualities for the existing approaches. In comparison, the proposed method shows a relatively better and more consistent result with lesser variation in the result qualities.\\
\indent Another observation is that, as FICA learns by token sequence comparison, it gets significantly slower as the overall size or the total number of tokens increases for a system. For example, considered lines of code for \lq Luaj\rq, were 36155 with an average of 79 tokens per clone fragment, resulting in the classification to take noticeably longer time than the proposed approach. We got the same behavior for similar relatively bigger software system like: \lq Java File Manager\rq and \lq Upm-swing\rq .

\begin{table*}
%% increase table row spacing, adjust to taste
\renewcommand{\arraystretch}{1.3}
% if using array.sty, it might be a good idea to tweak the value of
% \extrarowheight as needed to properly center the text within the cells
\caption{Comparison with Existing Systems}
\label{table:fica_comparison2}
\centering
%% Some packages, such as MDW tools, offer better commands for making tables
%% than the plain LaTeX2e tabular which is used here.
\resizebox{\textwidth}{!}{\begin{threeparttable}
\begin{tabular}{l|c|c|c|c|c|c|c}
\hline
\multirow{2}{*}{\textbf{Software System}} & %\multirow{2}{*}{\textbf{LoC clones}}\tnote{1} & \multirow{2}{*}{\textbf{Avg. Lines}}\tnote{2} & \multirow{2}{*}{\textbf{Avg. Tokens}}\tnote{2} &
\multicolumn{2}{c|}{\textbf{FICA}}  & \multicolumn{2}{c|}{\textbf{FICA Iterative}} & \multicolumn{2}{c}{\textbf{Proposed Method}} \\
\cline{2-7}
 &  \textbf{Precision} & \textbf{Recall}  & \textbf{Precision} & \textbf{Recall}  & \textbf{Precision} & \textbf{Recall} \\
\hline
\hline

Luaj  & 0.969642857 & 0.629930394 &	 0.97619047 &	0.769230769	
& 0.979827089	& 0.945319741	 \\
\hline
Ucdetector  & 0.951219512 &	\textbf{0.549295775} &	 0.971428571 &	\textbf{0.478873239} & 
0.895833333	&0.883561644 \\
\hline
Autocover Tool  & 0.830188679 &	0.956521739	&  0.843137255	& 0.934782609
& 0.926315789	& 0.967032967	\\
\hline
Upm-swing  & 0.989690722	& 0.738461538	 & 0.994923858	& 0.753846154	
& 0.985971944	& 0.944337812 \\
\hline
ipscan  & 0.863247863	& 0.918181818	
& 0.922330097	& 0.863636364	
& 0.964912281	& 0.800970874	\\
\hline
JavaGB  & 0.784722222	& 0.875968992	
& 0.792114695	& 0.856589147	
& \textbf{0.9}	&0.861878453	\\
\hline
JavaOcr  & 0.970588235	& 0.76744186	
& 0.973684211	& 0.860465116	
& 0.988304094	& 0.933701657	\\
\hline
JavaFileManager  & 0.962962963 & 0.882352941	
& 0.967254408	& 0.868778281	
& 0.941807044	& 0.725235849	\\
\hline
jMemorize  & 0.926829268	& 0.619565217	
& 0.933774834	& 0.658878505	
& 0.91576087	& 0.828009828	\\
\hline
FileBot  & 0.765217391	& 0.946236559	
& 0.791855204	& 0.940860215	
& \textbf{0.969581749}	& 0.676392573	\\
\hline
JAIMBot  & 0.993710692	& \textbf{0.619607843}	
& 0.98156682	& 0.835294118	
& 0.987980769	& 0.825301205	\\
\hline
JLipSync  & 1	& 0.517241379	
& 1	& 0.620689655 
& 1	& 0.857142857	\\
\hline

\end{tabular}

\begin{tablenotes}
    \item[1] Some of results are combination of detected clones from multiple clone detection tools (as listed in Table \ref{table:clone_detection_tools})
    \item[2] Average per code clone fragment
\end{tablenotes}
\end{threeparttable}}
\end{table*}

%%%%%%%%%%%%%%%%%%%%%%%%%%%%%%%%%%%%%%%%%%%%%%%%%%%%%%%%%%%%%%%%%%%%%%%%
%%% Evaluation with Different Code Clone Detection Tools
%%%%%%%%%%%%%%%%%%%%%%%%%%%%%%%%%%%%%%%%%%%%%%%%%%%%%%%%%%%%%%%%%%%%%%%%
%\subsection{Evaluation with Different Code Clone Detection Tools}
%As the proposed method works based on the clones reported by
%different clone detection tools, it is also important to evaluate its
%performance in conjunction with different clone detection tools.
%We collected the clone reports from five different clone detection
%tools based on open source projects. Besides, as the clone detection result quality might vary
%greatly based on the complexity or source code structure of a
%given software system, to generalize we selected 500 clone pairs
%detected by five clone detection tools (as mentioned above) from
%five different open source software systems (as mentioned
%above). Users reported 315 and 185 of them as true and false
%clones respectively on their manual validation. We then applied
%the proposed method to find out its validation performance. The
%obtained result was: True Positive (TP) =311, False Negative (FN)
%=4, True Negative (TN) =115 and False Positive (FP) =70. It is
%noticeable from the experimental result that it could successfully
%validate almost all the true clones with a precision of 0.82 in
%comparison to the average precision of the tools (0.63). This
%improvement can even be more useful for large-scale software
%systems.

%\section{FICA}

\section{Result Discussion}\label{sec:res_discusion}
\begin{figure}[tb]
\centerline{\includegraphics[scale=.6]{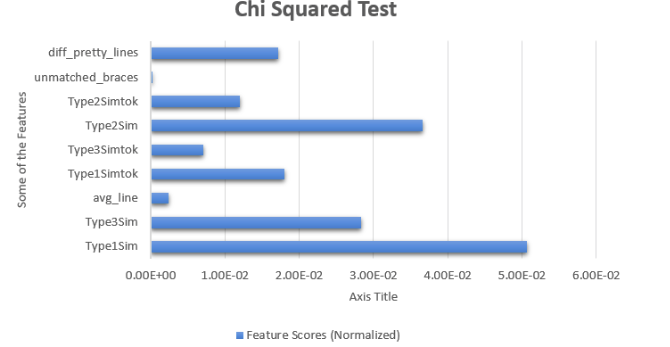}}
\caption{Feature Score via Chi-Squared Test}
\label{fig_chi}
\end{figure}

\begin{figure}[tb]
\centerline{\includegraphics[scale=.5]{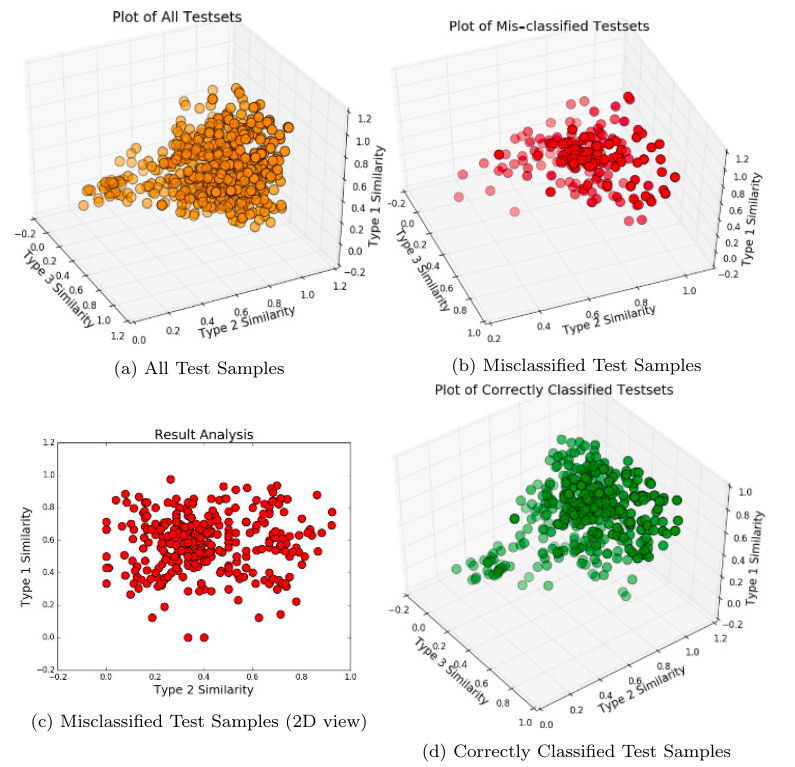}}
  \caption{Classification Result Analysis for Different Types of Code Clones}
   \label{fig_diff_clone_types}
\end{figure}

Artificial Neural Networks are efficient computing models which
can approximate complex functions. Different variants of Artificial Neural Networks have been widely
used for pattern recognition in different branches of computer
science \cite{cho1997neural},\cite{looney1997pattern},\cite{rocha2016artificial}. On the other hand, one of the major
criticisms is their being black boxes, since no satisfactory
explanation of their behavior has been offered. That is ANNs
are only given the inputs in the input layer and informed about
expected output from the output layer. ANNs then assign
required node biases and layer connection weights to predict
accordingly without providing us much information about the
complex function it learned or how it learned. So from the
perspective of our proposed method, it is challenging to know
the nature of the function the Neural Network has learned or if it
is giving its decision biasing completely on any of the features
used. \\
\indent However, assuming the Neural Network as a \lq black box' in
the middle of input sets and its predicted decision we tried to
find out if there is any bias on any feature of the Neural
Network on its output decision. Based on the classified test
samples by the algorithm we calculated feature contribution
scores using \textit{Chi Squared Test}. If the score is too high for a
particular feature in comparison to the rest, then it gives some idea about the Neural Network being biased to the
particular feature. Figure \ref{fig_chi}, shows the scores of some of the
selected features having higher scores out of all possible 
extracted features. From the figure, it is noticeable  that the normalized
score is kind of randomly distributed over the features rather
than being completely dominated by one or more features. This
gives us some idea that the trained model is not noticeably
biased toward any feature(s) on its decision making. Besides, the top 3
scores are found to be Type 1, Type 2 and Type 3 code clone
similarities respectively which is logical for the stated clone
validation problem. Similarly, average code clone fragment size or unmatched braces has much less contribution score as discussed in Section \ref{data_distribution}. The Chi-Squared test also supports these findings as noticeable from its low corresponding feature score in classification.\\
\indent Another important aspect to analyze from the proposed
method classification result is to see if it fails or succeeds only
for a particular type of clone(s). For example, it might be that the
model can only validate Type 1 clones and cannot validate the
other complex types of clones or there can also be possibility that
the proposed method fails to validate all the Type 1 clones. Especially, Type 3 clone is different and difficult to
validate in comparison to Type 1 or Type 2 code clones. That is
depending on the given type of code clone there is some
difference in the validation processes. This leads to the
possibility that any proposed method may work only with some
particular type(s) of clone(s). To analyze if there are any such
failure or success patterns for validation in the proposed method,
we plotted the classification result in 3D space where the axes
represent 3 different types of clones: Type 1, Type 2 and Type 3.
The plotted result is shown in Figure \ref{fig_diff_clone_types}. The top left plot of the
figure shows the scatter plot for the test samples along the 3 axes
each representing 3 different types of clone similarity. From the
plot, we can notice the test samples are randomly scattered in the
3D space representing the presence of all types of code clone
being available in the test samples. The top right plot of the
figure shows the scatter plot of the test samples that our
proposed method misclassified. The randomness of the scatter
plot suggests that the proposed method did not fail to classify
any particular type of code clones. For example, if the algorithm
would fail to correctly classify all the Type 3 clones then in the
scatter plot, all the misclassified test sample plot would more or
less aligned along a pariticular axis, such as \lq Type 3 Clone Similarity'. Besides the
bottom left plot of the figure shows a single plane (Type 1 vs
Type 2 plane) of the plotting for easier visualization. From this
plot, the randomness is clearly noticeable. From those studies
on the misclassified test samples by the proposed method, we can
 conclude that it did not fail for any particular type of clone. Similarly, the proposed method can successfully classify
all three types of code clones as we can notice from the
randomness of the correctly classified test samples in the bottom right
plot of the figure.\\

\indent Based on these findings we answer the research questions as
follows:\\
\indent \textbf{Answering RQ 1}, (Can the manual code clone validation
process be assisted via machine learning?): The proposed machine
learning algorithm was trained and tested via 10-fold cross
validation on a larger dataset. The
Neural Network converged within a range of 500 to 600 epochs.
Validation accuracy given by the proposed method is found to be
87.4\% after averaging each of the 10 folds testing. Besides the
trained system was tested by completely different 
software systems. We found the proposed method to come up
with promising accuracies for clone validation. These positive results reveal opportunities for
using machine learning for clone validation to assist in overall
code clone maintenance and analysis process.\\

\indent \textbf{Answering RQ 2}, (Does the proposed machine learning based
validation method works across different clone types and clone
detection tools?): To test the generality of validation, the
proposed machine learning method was tested with different
clone types and clone detection tools. From the evaluation study
we found that the method does not fail for any particular
type of clones. From the plotting of correctly classified or
validated clones by the proposed method in Figure \ref{fig_diff_clone_types}, it is
noticeable that the clones are randomly scattered across three
axes representing the validation works for all the three different
types of clones (as previously discussed). Similarly, the plotting
of misclassified clones also shows randomness across three
different axes. That indicates that the clone validation by the
proposed method does not succeed or fail for any specific types
of clone. These results demonstrate the
generality of the machine learning approach for working across
different types of clones. Besides, the method was evaluated on
validating the clone detection result by 5 different tools. As
shown in Figure \ref{fig_sys_acc}, the validation result in conjunction with
different tools found to be promising.

\section{Threats to Validity}\label{sec:threats_to_validity}
Neural Networks are widely used for modeling complex non-
linear relationships which traditional statistical methods often
fail to model accurately. However to learn such complex non-
linear functions Neural Networks need a larger training set and
also a good amount of time. So if this training phase is carried
out by individual programmers, the Neural Network might lack
 enough data, as it needs those clones to have been 
manually validated beforehand for training. Besides, even if the
individual programmers manage to have enough 
manually validated clones, the training process of the Neural
Network might take a significant amount of time, which might
reduce the usability of automatic clone validation. To make
this training phase easier, we validated a larger set of data by 5
different programmers and used them for training the Neural
Network. Though it removes the time-consuming step of
training for individual programmers, it might also
raise some threats to validity as the trained model is not based
on the individual programmer's choice at the beginning. However, as the model training was generalized by 5 different
programmers' independent decisions, this possible threat to
validity might be considered minor. Besides, to mitigate this
possible threat, the Neural Network weights are updated by
individual programmers feedback while being used. This way the
Neural Network converges towards the validation preferences of
the individual programmer while using it over time.\\
\indent The accuracy and precision of our work across different
software systems and clone detection tools was evaluated
against pre-judged true positive or false positive clones. These
judges can be affected by the subjective preferences on clones of
 individual programmers thus raising some possibility of
threats to the validation of the work. However, we tried to
mitigate this possible threat to validity by taking the validation
decision from multiple programmers.\\
\indent Another likely threat to validity is the possibility of having some
minor errors with feature extraction. For the extraction of used
features by the proposed method we had to use some source
code parsers that work via different transformations of the
source code. As the parsers are not always guaranteed to be
100\% perfect, the error (if any) might possibly propagate to the
feature calculation. However, best efforts were given to reduce
the probability of having any such errors in feature
calculations to make the evaluation as accurate as possible.

\section{Conclusion}\label{sec:conclusion}
In this paper, we introduced a machine learning based approach for
automatic code clone validation. Code Clone Detection tools
usually return the list of possible clones following some complex
searching procedure. The result often contains a large number of clones and often does not consider the preferences of
user's opinion or requirement. This leads to manual validation of
the result from the clone detection tools which gets worse
for large-scale software systems. We have proposed a machine
learning approach that assists in automatic validation of code
clones. The method takes feedback from the user to improve its
prediction on validation. We evaluated the proposed system with
different users, clone detection tools, artificially created code
clones, and open source projects. We found promising accuracy
with the automatic validation of clones by the proposed method.\\
%\indent While manually validating some code clones by ourselves and
%also surveying other programmers’ opinions we found that
%human judges in addition to finding out structural or syntactical
%similarity also tries to focus on understanding the contexts or
%functional similarity of the codes to decide if they are true
%positive or false positive clones. So our future work plan is
%finding out some of such features to further improve the result.

\bibliographystyle{unsrt}%Used BibTeX style is unsrt
\bibliography{main}

\end{document}